\documentclass[manuscript]{acmart}

\usepackage{multirow}
\usepackage{xurl}

\usepackage{tikz}

\newcommand\ChapterPrecis[2]{%
\begin{tikzpicture}[remember picture,overlay]
\node[anchor=north, draw=black, fill=yellow!20, inner sep=3pt, rounded corners, align=left, yshift=-#1] at (current page.north) 
{\parbox[t][1.3cm][l]{\textwidth}{\small #2}};
\end{tikzpicture}%
}

\AtBeginDocument{%
  \providecommand\BibTeX{{%
    \normalfont B\kern-0.5em{\scshape i\kern-0.25em b}\kern-0.8em\TeX}}}

\setcopyright{acmcopyright}
\acmJournal{TOIT}
\acmYear{2022} \acmVolume{1} \acmNumber{1} \acmArticle{1} \acmMonth{1} \acmPrice{15.00}\acmDOI{10.1145/3561300}

\acmBooktitle{ACM Transactions on Internet Technology}
\begin{document}

%%%%% Arxiv Reference
    \ChapterPrecis{1cm}{If you cite this paper, please use the ACM Transactions on Internet Technology reference: Massimo La Morgia, Alessandro Mei, Francesco Sassi, and Julinda Stefa.
2023. The Doge of Wall Street: Analysis and Detection of Pump and Dump
Cryptocurrency Manipulations. \textit{ACM Trans. Internet Technol.} 23, 1, Article
11 (feb 2023), 28 pages. \color{blue}{\url{https://doi.org/10.1145/3561300}}
}
%%%%

\title{The Doge of Wall Street: Analysis and Detection of Pump and Dump Cryptocurrency Manipulations}


\author{Massimo La Morgia}
\affiliation{
  \institution{Sapienza University of Rome}
  \country{Italy}
}
\email{lamorgia@di.uniroma1.it}

\author{Alessandro Mei}
\affiliation{
  \institution{Sapienza University of Rome}
  \country{Italy}
}
\email{mei@di.uniroma1.it}

\author{Francesco Sassi}
\affiliation{
  \institution{Sapienza University of Rome}
  \country{Italy}
}
\email{sassi@di.uniroma1.it}

\author{Julinda Stefa}
\affiliation{
  \institution{Sapienza University of Rome}
  \country{Italy}
}
\email{stefa@di.uniroma1.it}

\renewcommand{\shortauthors}{La Morgia, et al.}

\begin{abstract}

Cryptocurrencies are increasingly popular. Even people who are not experts have started to invest in these assets, and nowadays, cryptocurrency exchanges process transactions for over 100 billion US dollars per month. Despite this, many cryptocurrencies have low liquidity and are highly prone to market manipulation. 
This paper performs an in-depth analysis of two market manipulations organized by communities over the Internet: The pump and dump and the crowd pump.
The pump and dump scheme is a fraud as old as the stock market. Now, it got new vitality in the loosely regulated market of cryptocurrencies. Groups of highly coordinated people systematically arrange this scam, usually on Telegram and Discord.
We monitored these groups for more than 3 years detecting around 900 individual events. We report on three case studies related to pump and dump groups. 
We leverage our unique dataset of the verified pump and dumps to build a machine learning model able to detect a pump and dump in 25 seconds from the moment it starts, achieving the results of 94.5\% of F1-score.
Then, we move on to the crowd pump, a new phenomenon that hit the news in the first months of 2021, when a Reddit community inflates the price of the GameStop stocks (GME) by over 1,900\% on Wall Street, the world's largest stock exchange. Later, other Reddit communities replicate the operation on the cryptocurrency markets. The targets were DogeCoin (DOGE) and Ripple (XRP). We reconstruct how these operations developed and discuss differences and analogies with the standard pump and dump. 
We believe this study helps understand a widespread phenomenon affecting cryptocurrency markets. The detection algorithms we develop effectively detect these events in real-time and help investors stay out of the market when these frauds are in action.
\end{abstract}




\maketitle

\section{Introduction} 

Pump and dump is a market manipulation fraud that consists in artificially inflating the price of an owned security and then selling it at a much higher price to other investors~\cite{kyle2008define,kramer2005way}. This fraud is as old as the stock market. One of the most famous pump and dumps in Wall Street history happened in the late '20. The security was the RCA Corporation. RCA was the manufacturer of the first all-electric phonograph, one of the hottest pieces of technology at that time. The fraud was organized by the "Radio Pool", a group of investors that artificially pumped RCA to the incredible price of \$549, and then dumped the shares making the price plummet to under \$10. A large number of investors lost all of their savings in this operation.
At that time, communication was done through the radio, tabloids, and word of mouth.

With the advent of the hectic and almost non-regulated markets of cryptocurrencies, pump and dumps are more vital than ever.
There are now hundreds of cryptocurrencies, the market is not strictly regulated, and prices are easy to manipulate. Thus, pump and dump schemes are incredibly common, with public groups on the Internet, rules, and precise and complex organization.
Now, pump and dumps are led by a large number of self-organized groups over the Internet, and the phenomenon is viral though still not very well known.
As of January 2021, a new kind of operation has been in the global spotlight. A group of people active on a Reddit group called r\textbackslash{wallstreetbets} started an operation against a few hedge funds shorting GameStop stocks (GME). The group was able to attract other people and managed to raise the stock price of GME by more than 1,900\%~\cite{redditGME}.
The event got worldwide attention, and several celebrities, including Elon Musk, rock star Gene Simmons, and rapper Snoop Dogg~\cite{markcuban} commented on the event and contributed to making it even more popular.
Following the success of this operation, people collaborated into buying other stocks such as AMC (AMC Entertainment Holdings), BB (BlackBerry Ltd.), and NIO (NIO Inc.) and later the Ripple (XRP) and DogeCoin (DOGE) cryptocurrencies. Also in these cases, prices increase rapidly in a few days~\cite{otherstocks}.

In this work, we describe the pump and dump phenomenon in the cryptocurrency ecosystem, focusing on the organization of the groups and the frauds. 
We perform a 3 years longitudinal analysis of the pump and dump operations on 4 different exchanges. Then, we analyze the events arranged by Big Pump Signal, a pump and dump group that moved 5,176 BTC (around \$300M as of today) in a single operation.
Lastly, we introduce a novel detection algorithm that works in real-time.
The algorithm is not just based on the detection of the abrupt rise of the price. The fundamental idea is to leverage the abnormal growth of so-called \emph{market buy orders}, buy orders that are used when the investor wants to buy extremely quickly and whatever is the price. Just like the colluding members of a pump and dump group when the pump starts.
Moreover, we describe a new kind of pump operation---that we refer to as crowd pump to distinguish it from the standard pump and dump, discussing the differences in the organization and aim between the standard pump and dump and the crowd pump.

Our main contribution are:
\begin{itemize}
    \item \textbf{Pump and dump dataset.} We publicly released our dataset~\cite{pumpdataset} containing more than 1,000 confirmed pump and dump events arranged by 20 different Telegram groups. 
    \item \textbf{Pump and dump detection model.} We propose a novel real-time machine learning model, showing that it outperforms the current state of the art~\cite{kamps2018moon}, improving the expected speed of the detection from 30 minutes to 25 seconds and, at the same time, the F1-score from 62.7\% to 94.5\%.
    \item \textbf{Crowd pump analysis.} We conduct an in-depth analysis of the crowd pump events carried out on the DogeCoin and Ripple cryptocurrencies. Collecting and analyzing the messages on Reddit, we reconstruct the way these events occurred and how they started. Lastly, we show that it is possible to use the proposed machine learning model to detect when a crowd pump is in action.
\end{itemize}

\section{Pump and dump groups}
\label{sec:pump_and_dump_groups}
Pump and dump schemes are performed by self-organized groups of people over the Internet. These groups arrange the frauds out in the open on the Telegram~\cite{Telegram} instant messaging platform or Discord server~\cite{Discord}. Thus everyone can join the groups without prior authorization. 
Along our longitudinal research, from July 2017 to January 2021, we joined and followed all the activities performed by more than 100 groups daily. Being members of the groups allowed us to retrieve and collect one-of-a-kind information such as internal group organization, the phases of pump and dump arrangement, and how the groups attract outside investors inside the market.
In the following section, we report on the findings we discovered about these communities.

\subsection{Group organization}
\begin{figure}
\includegraphics[width=0.7\textwidth]{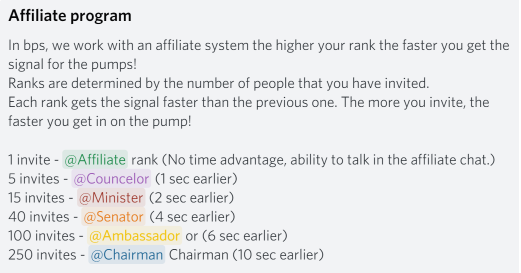}
\caption{Affiliate program and benefits of the Big Pump Signal group.}
\label{fig:discord}
\end{figure}

Pump and dump groups have leaders (or admins) that administrate the group, and a hierarchy of members. If a member is higher in the hierarchy, he gets the message that starts the pump by revealing the target cryptocurrency a few moments earlier than lower ranked people. This way, the member has a higher probability of buying at a lower price and make more money from the pump and dump operation. The advantage in terms of time of being at a higher level is usually between 0.5 and 1 second with respect to the next level, and the maximum advantage is in the interval between 1 and 10 seconds.  Most groups are organized as an affiliation system ---climbing the hierarchy is possible by bringing new people into the group. The larger is the number of new members brought to the group, the higher the ranking. Fig.~\ref{fig:discord} shows the affiliation system of the Big Pump Signal group and the rank's benefits.

Some groups have a simple hierarchy with only two levels: Common members and VIP members. In these groups, to become a VIP the user has to pay a fee, usually in Bitcoins, in the range of 0.01 to 0.1 Bitcoins (from approximately \$310 to \$3,100 at current exchange rates\footnote{Data retrieved on January 10, 2021}). 
In the pump and dump groups, the admins are the only people that make decisions. We saw only in rare cases the admins running polls to decide the hour of the pump or the exchange to use but never to decide the target cryptocurrency.

\subsection{Group communication}

The groups typically use Discord servers and Telegram channels to communicate and organize the pump. Telegram~\cite{Telegram} is an instant messaging service, and a Telegram channel is a special kind of chat in which only the owner of the channel can broadcast public messages to all the members. Discord~\cite{Discord} is a VoIP and text chat service. It was originally designed for video gaming communities, but nowadays it is widely used by communities not related to video games~\cite{la2021uncovering}. 
Discord offers the possibility to create macro sections and host multiple chat rooms. Each section has its own topic or scope. In our analysis, we have found that all the pump and dump Discord servers are organized in roughly the same fashion, with the following sections: 
\begin{itemize}
  \item \textbf{Info \& How-Tos}: These two sections are like an electronic bulletin board with pinned messages. Both sections are composed of several rooms that contain only one or very few messages. The rooms of the Info section usually contain the rules of the group, the news about the group, how the affiliation system works (Fig.~\ref{fig:discord}), and the F.A.Q.. The rooms of the How-Tos section contain manuals related to the cryptocurrency world or the best practices to participate in a pump and dump operation.
  \item \textbf{Invite}: This section contains rooms where the bots of the server live. Here, the users can query the bots to generate invite links to bring new members or to know the number of people that joined the server by using their invite links.
  \item \textbf{Signal}: This is the core section of the group, in which only the admins can write. Usually, there are two rooms in this section: The pump-signal and the trading-signal. In the first room, the admins share info about the next pump and dump operation. In the second, they share trading advice.
  \item \textbf{Discussion}: In this section, there are rooms covering different topics where the group members can freely chat.
\end{itemize}
Usually, the messages written in the news and in the pump-signal rooms are also broadcasted to the Telegram channel.

\subsection{Organization of the pump and dump operations}

\begin{figure}
\includegraphics[width=0.8\textwidth]{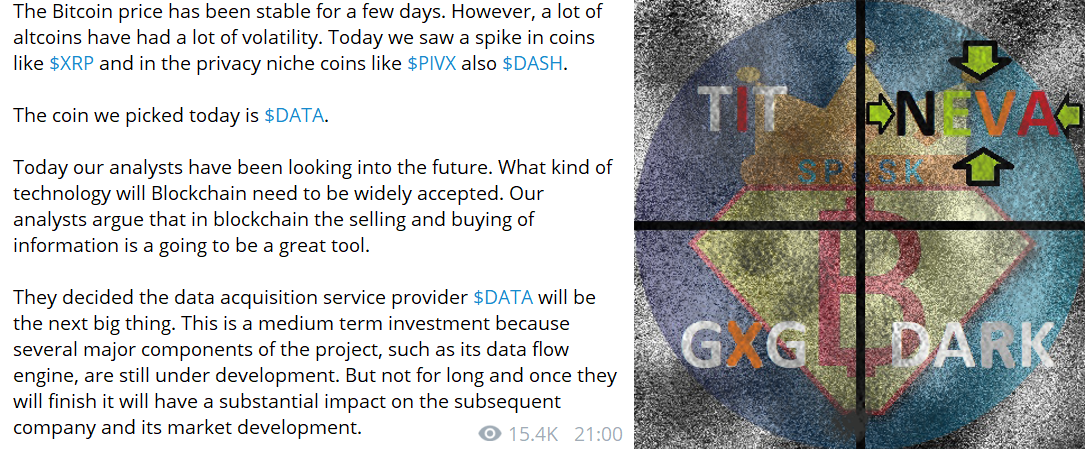}
\caption{Messages that indicate the start of a pump and dump operation on the Streamr DATAcoin (on the left) and the NevaCoin (on the right).}
\label{fig:coin_signal}
\end{figure}

The levels of activity of the many pump and dump groups on the Internet differ considerably. The most active ones perform roughly one pump and dump operation a day. Less active groups perform one operation a week. Other groups perform operations only when they believe the market conditions are good. The steps during the operation are typically as follows:
\begin{itemize}
\item A few days or hours before the operation the admins announce that the pump and dump will happen and communicate which is the exchange that will be used, the exact starting time of the operation, and whether the operation will be FFA (Free for All---everybody gets the message at the same time) or Ranked (VIPs and members of higher levels in the hierarchy get the starting message before the other members).
\item The announcement is repeated several times, more frequently as the starting time of the operation gets closer.
\item A few minutes before the start, the admins share some simple tips and best practices: Check your Internet connection, buy low and sell high, disconnect all the other Internet activities to get low latency on your network, hold the currency as much as possible waiting for an external investor. At this point, the free chat rooms are closed in order to avoid so-called FUD (Fear, Uncertainty and Doubt)---sometimes due to actual human anxiety of losing money, sometimes due to activities of disinformation done by people that have the goal of sabotaging the operation, make people panic and make the panic spread in the group. This is also useful to avoid any possible overload on the communication server.
\item At the established time the targeted cryptocurrency is revealed, the exact time depends on the position in the hierarchy of the group. Usually, the name of the cryptocurrency is contained in an image that is obfuscated in such a way that only humans can read it correctly. Fig.~\ref{fig:coin_signal} shows an example, a message that instructs to start a pump and dump operation on the NevaCoin. The idea behind the obfuscation is to make it hard for bots to parse the message with OCR techniques and start the operation faster than humans. 
\item A few seconds after the start of the operation, the admins share a piece of news and invite all the group members to spread the information that the price of the cryptocurrency is rising. This is done in dedicated chat boxes, forums, and Twitter. This activity aims to attract external investors by creating FOMO---Fear of Missing Out a unique investment opportunity.

\item Finally, when the operation ends, the admins reopen the free chat rooms and share some statistics about the pump with the members.

\end{itemize}

\section{Case study}
In this section, we present three case studies. In the first, we perform an analysis of the pump and dump groups, the targeted exchange, and the cryptocurrencies. In the second, we focus on Big Pump Signal, arguably the biggest pump and dump group, able to generate a volume of transactions of 5,176 BTC in a single operation. Lastly, we present the case study of the Yobit exchange that organized 3 pump and dump operations in 2018. We analyze these frauds and leverage the users' comments on Twitter regarding these events to understand the feeling of the crypto-community about the phenomenon.

\begin{table*}
	\centering
	\small
    \caption{Metrics of pump and dump groups.}
    \label{tab:group_users}
    \begin{tabular}{l r r r r r r r}
        \toprule
        Group name & Telegram Users &Discord Users & Hierarchy & Main Exchange & PnD (\#) & avg. Volume (\$)\\
        \midrule
        BigPumpSignal & $72,097$ & 104,830 & affiliation & Binance & 41 & 7,245,437 \\
        Trading Crypto Guide & $91,725$ & --- & vip & Binance & 22 & 2,442,923 \\
        Crypto Coin B & $166,689$ & --- & vip & Binance & 12 & 5,733,637 \\
        Crypto4Pumps & $11,716$ & --- & vip & Bittrex & 45 & 491,395 \\
        Pump King Community & $7,771$ & --- & vip & Bittrex & 14 & 931,960 \\
        Luxurious & $6,020$ & --- & free & YoBit & 17 & 4,997 \\
        AltTheWay & $7,333$ & --- & free & YoBit & 253 & 700 \\
        \bottomrule
    \end{tabular}
    
\end{table*}

\begin{figure}
    \includegraphics[width=1\textwidth]{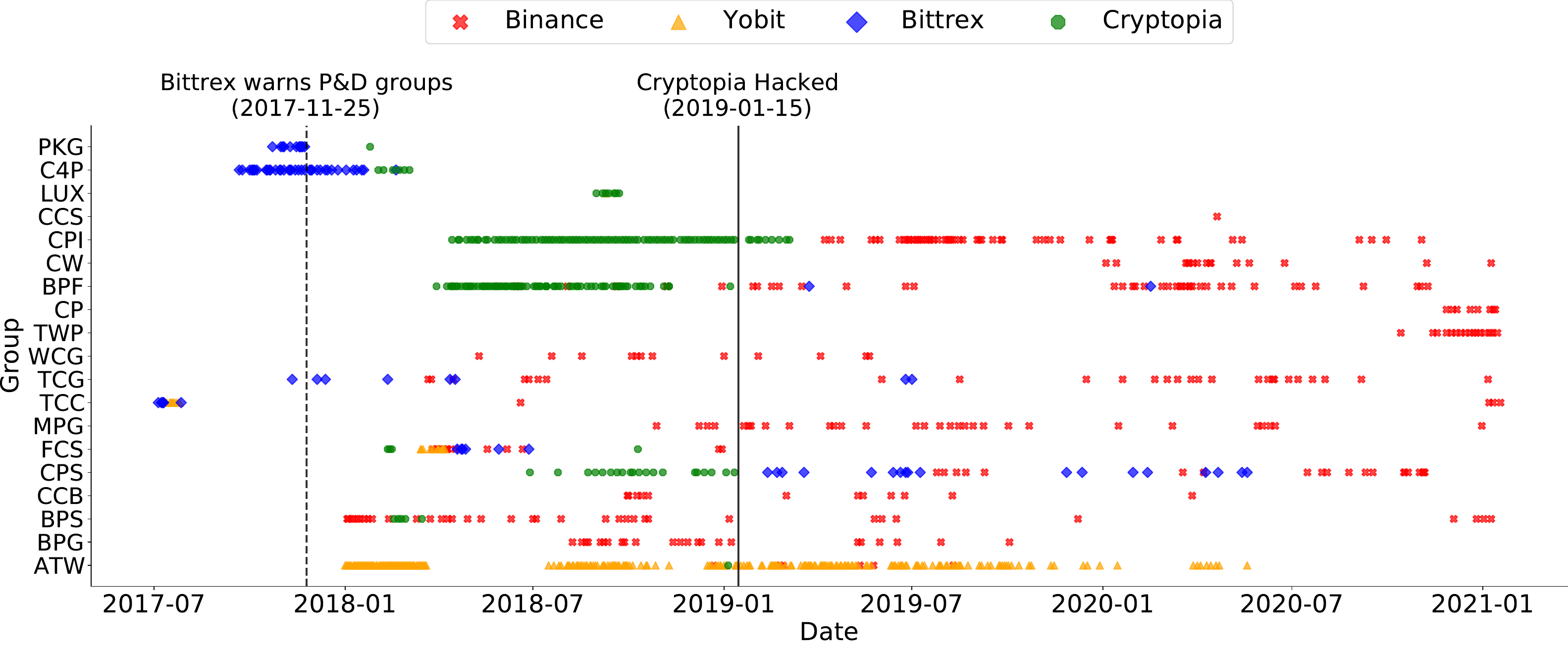}
    \caption{Pump and dump events by group and exchange during the period of the analysis.}
    \label{fig:historical_pumps}
\end{figure}

\subsection{The groups, the exchanges, and the target cryptocurrencies}

We conduct an in-depth investigation of the cryptocurrencies and the exchanges used for the pump and dumps. We do so in a period that goes from July 2017 to January 2021. In this period, we found more than 100 groups, by keywords search (e.g.: \textit{"Pump", "Dump", "Signal"}) on Telegram, Twitter, Reddit, BitcoinTalk~\cite{bitcointalk}, or manually extracting information from CoinDetect~\cite{coindetect} or the PADL~\cite{padl} Android app. From this set, we select 20 different groups since the others are not very active or have a small number of users. We include the list of the selected groups in the Appendix~\ref{ssec:monitoredgroups}. We report the extended name, their Telegram link, and the short version of the name that we will use in the charts for each group.

Table~\ref{tab:group_users} shows a few metrics about different kind of groups with respect to hierarchy, number of users, and number of pump and dump operations\footnote{Data retrieved on October 2018}.
Reading the Telegram channel history of these 20 groups, we found evidence of 1,108 pump and dump events carried out on 4 exchanges. We discovered that 206 of these operations were jointly arranged by more than one group. Hence we have in our dataset 902 unique pump and dump operations.
Analyzing our data, we found that the scheme involved 378 different cryptocurrencies,
only 340 of which CoinGecko still lists. CoinGecko is a service that exposes APIs~\cite{coingeckoapi} to retrieve historical trading data. Instead, leveraging the CryptoCompare API~\cite{cryptocompareapi} we were also able to retrieve the market capitalization, at the time of the fraud, for 264 coins. 
Analyzing the volume of the 24 hours before the pump of the target cryptocurrencies, we can see that 284 (83.5\%) of them moved less than \$1 million in total in all the exchanges. 182 (53.5\%) of them moved a negligible amount of money, less than \$10,000.
Also, analyzing the market capitalization of 264 coins, we find out that 140 (71\%) coins are below the \$20 million of market capitalization, with 44 (22\%) below \$1 million.

The market capitalization of targeted coins is low, considering that the first asset with less than \$20 million is at the $616th$ position of the cryptocurrency ranking by market capitalization\footnote{According to CoinMarketCap data retrieved on February 18, 2021}. Typically, Binance is the market of choice for the pump and dump operations on currencies with higher market capitalization, Cryptopia for those with lower market capitalization.
In particular, the median market capitalization of the cryptocurrencies for exchange is \$25,574,192 for Binance, \$2,619,703 for YoBit, \$2,512,627 for BitTrex, and \$144,373 for Cryptopia.
Thus, the target cryptocurrencies of pump and dumps have a very low net worth value and a vast circulating supply. 
Lastly, we find that $264$ (78.3\%) assets
are priced below \$$0.4$. As such, with a relatively small investment, pump and dump groups can buy huge amounts of cryptocurrencies and easily increase their price in the pump phase of the fraud.

Figure~\ref{fig:historical_pumps} shows the number of the pump and dump operations by group and exchange. 
The figure shows that the groups typically work on one or a couple of exchanges. That is quite normal. Indeed, if the groups jump from one exchange to the other, the members would be forced to move their assets according to the selected exchange, pay the withdrawal and the network fees, and waste their time.
Sometimes, the groups move from one exchange to another due to external circumstances.
For instance, 2 out of 3 groups that operated mainly on Cryptopia suddenly changed the target exchange after Cryptopia was hacked in January 2019. In contrast, the third group waits almost a month before moving to Binance. Another example is when Bittrex warned the community about its intention to ban users involved in pump and dump operations~\cite{bittrex}. We can note that before the warning (the dashed line on the figure), 42 out of 54 (77.8\%) operations are arranged on BitTrex. After, only 48 out of 817 (5.9\%).
From a longitudinal perspective, it is possible to note that, until May 2019, pump and dumps are evenly distributed among the four exchanges. After this date, Binance has become the most popular exchange among the groups.

\begin{figure}
        \centering
        \includegraphics[width=1\textwidth]{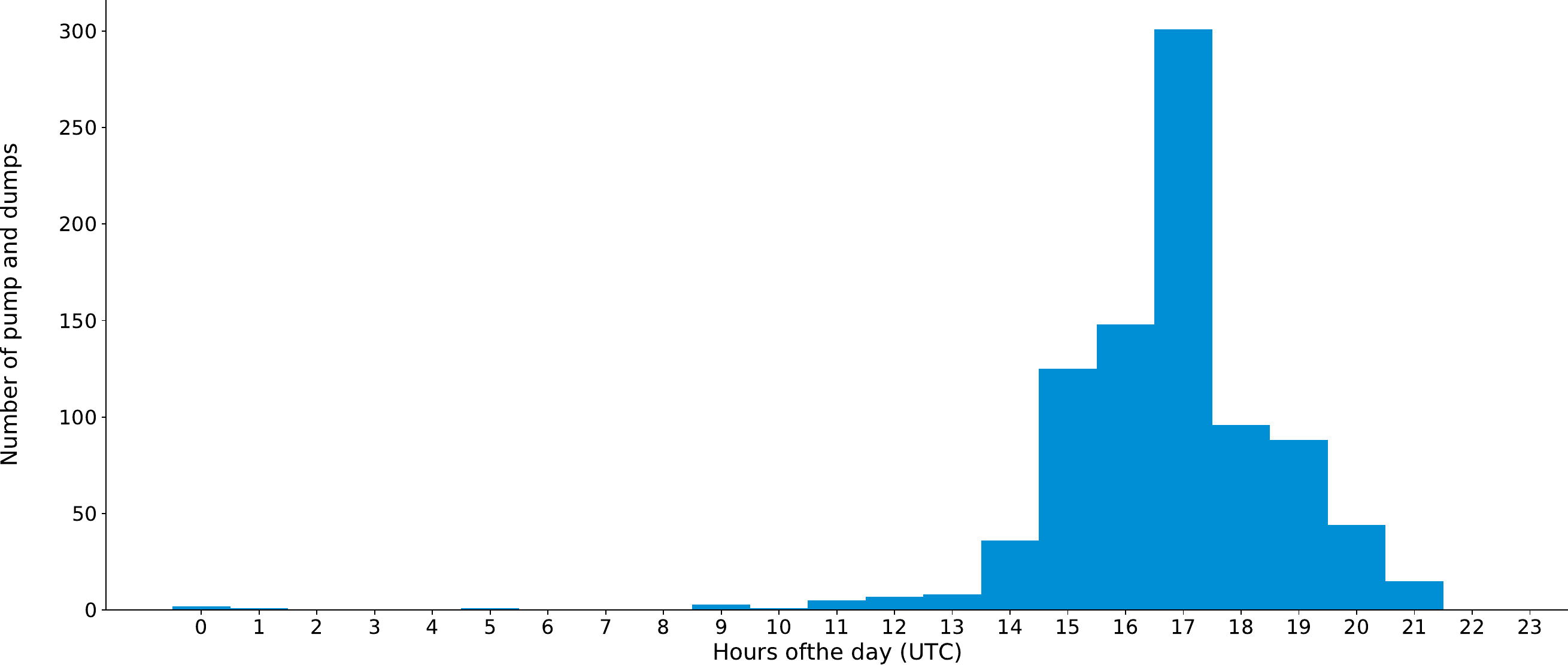}
        \caption{Pump and dumps during the hours of the day.}
        \label{fig:hours}
\end{figure}
Looking at Figure~\ref{fig:hours}, we can see that most of the pump and dump events are organized in the late afternoon of the European time-zones. Hours in which European web users are more active, according to~\cite{lamorgiadark,gill2007youtube}. Moreover, the Binance exchanges do not allow US citizens to use their service. This information could indicate that the admins and the members of the groups under investigation are mostly Europeans.

\subsection{YoBit}
YoBit is a Russian exchange active since August 2014. In October 2018, it processed almost \$1 billion, and it was the 43rd exchange by monthly traded volume. In October 10th, 2018, YoBit announced on Twitter that it is arranging a pump and dump event. More in detail, they claimed that they would buy 10 Bitcoins of a random coin, in a range of 10 minutes. After the first pump, done on the Putin Coin, YoBit repeated the event twice---on October 15th on the Lambo Coin and on October 17th on Chat. All three cryptocurrencies had practically no transactions---in the 24 hours before the pump the three coins moved \$36, \$800, and \$59 respectively. Table~\ref{tab:yobit_pump} shows the volumes and the prices during the events. The price of the Putin Coin, for example, has reached a peak of 14 times the opening price. The huge volumes and the high prices of the coins during the events went back to their original state a few hours later.

\begin{table*}
\small
\centering
\caption{YoBit pump and dumps.}
\label{tab:yobit_pump}
\begin{tabular}{l r r r r r}
\toprule
Cryptocurrency &  Date & Volume(\$) & Open price(\$) &Max price(\$) & Price increase \\
\midrule
Putin Coin  & 2018-10-10 & 955,077 &  0.0075  & 0.1131 & 1408\% \\
Lambo Coin  & 2018-10-15 &980,645 & 158.08 & 320,000 & 2024\%\\
Chat  & 2018-10-17 & 661,109 & 0.0320 & 0.3839 & 1099\%\\

\bottomrule
\end{tabular}
\end{table*}

This unprecedented behavior of an exchange hit the news~\cite{yobitscam0,yobitscam1,yobitscam2} and the community started to tweet about it. 
We collected all the tweets sent as a reply to the announcements by YoBit to see the impact that the event had on the community and their feelings about pump and dump schemes. We got 517 tweets, among which we removed 46 tweets containing images only and 157 tweets not related to the pump and dump events. We analyzed the remaining 314 tweets with the Google Cloud Natural Language API~\cite{googleapi} to get the sentimental score on the reaction of the community. After the analysis, we got that 46.5\% of the tweets had a negative sentiment on the events. 42\% a neutral feeling, and only 11.5\% of the tweets a positive feeling. 
Moreover, several crypto-influencers on Twitter strongly commented against YoBit, such as Rudy Bouwman, co-founder of DigiByte, the 37th cryptocurrency by market capitalization at the time.

\subsection{The Big Pump Signal group}
\begin{figure}
  \includegraphics[width=1\textwidth]{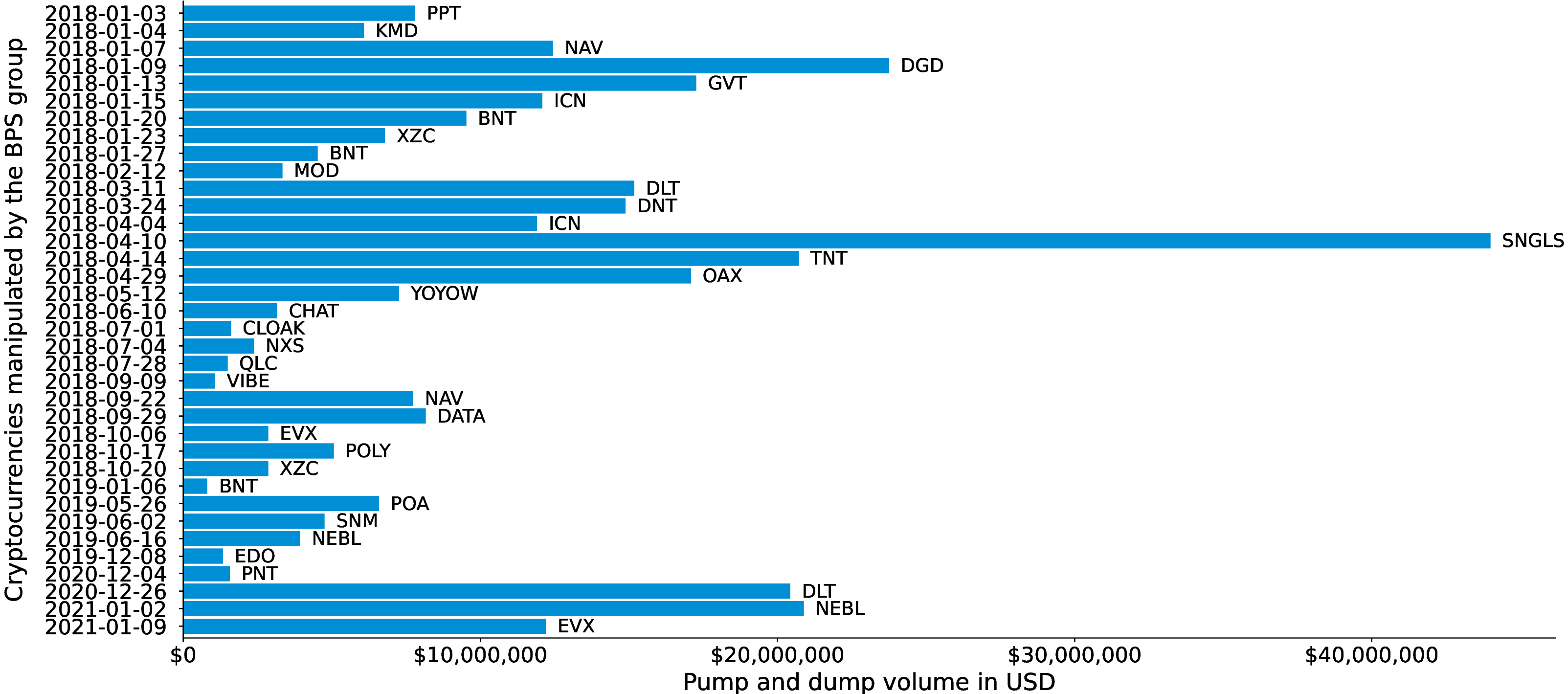}
  \caption{Big Pump Signal pump and dump operations.}
  \label{fig:BPS_pumps}
\end{figure}

With a peak of more than 200,000 members on Telegram and 250,000 members on Discord in January 2018, Big Pump Signal (BPS) is arguably the largest pump and dump public community on the Internet. Reading the pump announcements on the Telegram channel of Big Pump Signal, we found 41 pump events organized by them, 36 of which carried out on the Binance exchange and 5 on Cryptopia. 
Figure~\ref{fig:BPS_pumps} shows the operations carried out by the group and their volume.
Throughout all their pump and dump operations, the group moved globally \$129,674,881 within an average of 5 minutes from their start; including the time interval of the whole operations, the group moved globally \$343,433,660. Their most successful pump and dump was arranged on May 10, 2018, when they targeted the SingularDTV (SNGLS) alt-coin. In this operation, the value of the SNGLS coin sharply oscillated for more than 9 hours and recorded a volume of around 5,176 Bitcoins (\$36,750,000).

BPS has an affiliation hierarchy. The highest level is achievable after inviting $250$ new members. In ranked pump and dump operations, the affiliation guarantees the members to receive the signal between $1$ and $10$ seconds before the unranked members. 
Since the beginning, the Big Pump Signalers have promoted their group by advertising on social networks like Twitter and Quora. Thanks to their aggressive marketing campaigns and the hype on cryptocurrencies in late 2017, the Big Pump Signal group has grown extremely fast.

As the group grew larger, the admins started also targeting cryptocurrencies with medium market capitalization. The admins claim that they base the choice of the coin on technical analysis. They also claim to re-pump the cryptocurrencies by collaborating with a small investment firm. The investment firm is believed to be frequently involved in organizing pump and dumps on their own. Examples are the pump and dumps of the Monetha coin (MTH) and the WePower (WPR) coin on the Binance platform on September 17, 2018.
Our analysis shows that BPS typically chooses cryptocurrencies with a steady price and news coverage in the recent past. They leverage the news coverage to generate interest and attract external investors. An example is the retweets of news from the fake Twitter account of John McAfee (e.g. @oficiallmcafee, and @TheJohnMcafee) belonging to the admins of the group. 

\begin{figure}
  \includegraphics[width=1\textwidth]{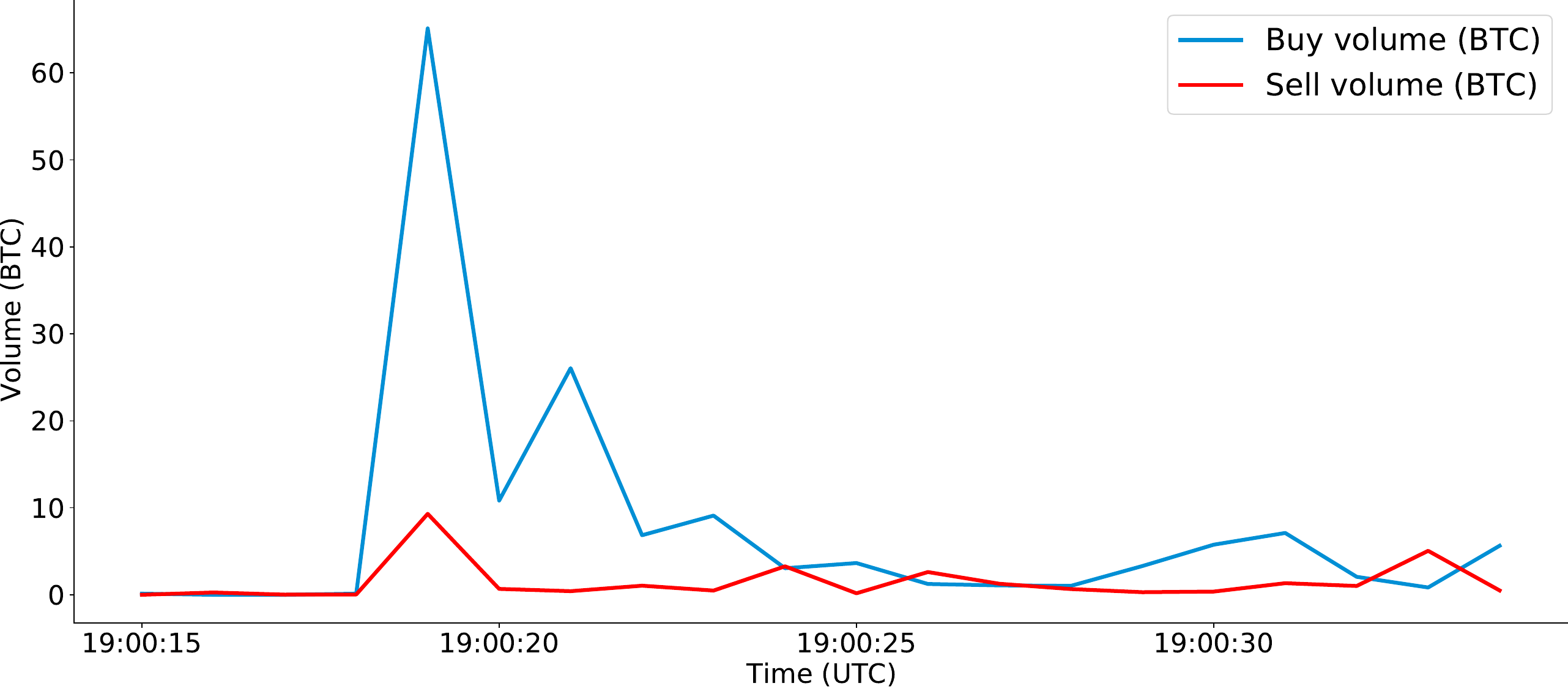}
  \caption{Pump and dump on the OAX coin.}
  \label{fig:BPS_oax}
\end{figure}

\subsubsection{Analysis of the BPS pump phase}
The BPS group moves large volumes of Bitcoins in each operation. Figure.~\ref{fig:BPS_oax} represents a zoomed image of the very first 30 seconds of the pump on the OAX cryptocurrency. The blue line in the figure represents the volume of buys; the orange line the volume of sells. We observe that the buy and sell volume in the first seconds is very close to zero. Then, there are two buy peaks (blue line in Figure~\ref{fig:BPS_oax}) of approximately 65 Bitcoins (sec. 19) and 26 Bitcoins (sec. 21) respectively. The two peaks correspond to the actions of VIPs and the common members---a normal behavior, considering that the group has a ranked policy. We also observe a peak in the sell volumes (orange line in Figure~\ref{fig:BPS_oax}) of almost 10 Bitcoins at the moment of the first buy peak, the 19th second. Considering that group members are still buying and the reaction time for outsiders is too short, this sudden big sell volume is abnormal. There can be only two possible actors to sell their assets: the bots and the admins. To discern between the two, we need to investigate the single transactions. Our analysis shows that, as the price rises, there are many small sell operations at incremental values, probably by the arbitrage bots. Then, we observe a last single shot transaction for over $4$ Bitcoins when the OAX coin reaches the trading value of 0.00012 BTC, probably done by the admins of the group. We believe they have operated through a \textit{sell limit} trade order---a conditional order triggered when the price of a trading pair reaches/out-tops a given value. Of course, the same order could have also been placed by an outside investor. However, we believe that a sell limit of that amount, 41\% more than the initial price, is most likely due to an insider.

\section{Pump and dump detection}

\subsection{The idea}
As we know, standard investors are the victims of pump and dump schemes. When they see that the price of a cryptocurrency rises, they can believe it can be a good investment opportunity. This is not the case when a pump and dump scheme is in action---the rise does not have economic grounds. It is just market manipulation. In order to protect investors, it is crucial to understand if we can detect a pump and dump in action and how quickly. This is the goal of this section.

To better understand how we can detect pump and dumps, it is essential to have some basic notions. The pending orders for a cryptocurrency, like securities, are listed in the \emph{order book} for that cryptocurrency. The book is a double sorted list of sell (ask) and buy (bid) orders not yet filled. The asks are sorted from the lowest price to the highest, the bids from the highest to the lowest.
The fastest way to buy on the market is through a \emph{buy market order}. A buy market order looks up the order book and fills all the pending asks until the requested amount of currency is traded. Although a market order completes almost instantly, the price difference between the first and the last ask needed to fill the order can be very high, especially in markets with low liquidity. So, the total cost of the order can be unpredictably high. A more careful investor would use \emph{limit buy orders}, orders to buy a security at no more than a specific price. Buy market orders are not frequent in everyday transactions, and investors use them when they need fast execution, just like the members of pump and dump groups in action. Our idea is to use this pattern and other information about volume and price to detect when a pump and dump scheme starts.

\subsection{The data}
\label{sec:dataset}

As highlighted by Kamps et al.~\cite{kamps2018moon}, it does not exist a dataset in the literature of the confirmed pump and dumps. Thus, we need to build one for this work. From the 20 groups we joined, we selected only the pump and dump schemes carried out on Binance. We made this choice for two main reasons. The first one is that Binance exposes APIs~\cite{binanceapi} that allow retrieving every single transaction in the whole history of a trading pair differently from other exchanges. The second is that pump and dumps on other markets are usually carried out by groups with few active members and economic resources. These groups can only target alt-coins that have almost no volume of transactions for days before the scheme. Thus, we believe that pump and dumps carried out on Binance are the most interesting and challenging to detect.

From the initial set of pump and dumps, we select all the events on Binance---317 pump and dump events. We retrieved the historical trading data for each pump and dump for 14 days, seven days before and seven after the event. Some pump and dump are a few days apart on the same alt-coin, so we discarded duplicate days. In the end, we have globally about 900 days of trading. The data are a list of trade records: Volume, price, operation type (buy or sell), and the UNIX timestamps. The records belonging to the same order at the same price have aggregated quantities, and a single order filled at different prices is split into more records.

Unfortunately, the Binance APIs do not tell the kind of order (e.g.: \textit{Market, Limit, Stop Loss}) placed by the buyer, so we need to infer this information.
To do this, we can use the fact that market orders complete instantly, and we can aggregate the buy operations filled at the exact millisecond as a single market order. Since we do not know the original nature of these orders, we define them as \textit{rush orders}. A problem with this inference method is that it misses the market orders that are filled by the first ask of the order book. Still, we believe we have a good witness of market orders' abrupt rise even with this approximation. As a contribution to the community, we will publicly release this dataset\cite{pumpdataset}.

\subsection{Features and classifiers}
\label{sec:features}
To detect the start of the fraudulent scheme, we analyze several kinds of features. Then, we use them to feed two different classifiers: Random Forest and AdaBoost.
Random Forest~\cite{breiman2001random} is an ensemble learning method consisting of a collection of decision tree classifiers such that each tree depends on the values of a random vector sampled independently, each tree casts a vote, and the prediction is the most popular class between all the votes. 
AdaBoost~\cite{freund1997decision} is a meta-estimator that ensembles multiple weak classifiers---a classifier that performs slightly better than a random guess into a stronger one.
It starts by training a weak classifier that assigns the same weight to all the dataset instances. It then fits additional copies of the classifier on the same data, tuning the weights in favor of the previously misclassified instances.
In our case, the weak classifier is a Decision Tree~\cite{safavian1991survey} with a maximum depth of 5.
We built our features upon the idea of~\cite{siris2004application} for the detection of Denial of Service attacks through an adaptive threshold. Since we do not want to find a threshold in our case, we rework their idea in this way: We split data in chunks of $s$ seconds, and we define a moving window of size $w$ hours.

We conduct several experiments with different sets of features and settings regarding the window and the chunk sizes. Since our goal was to build a classifier that detect a pump and dump scheme as soon as possible from the moment it starts, the chunk size must be reasonably short. At the end of our study, we achieved the best F1-score with a chunk size of 25 seconds and a window size of 7 hours; and the best speed with a chunk size of 5 seconds and a window size of 35 minutes.
Here are the features we used:
\begin{itemize}
  \item \textbf{StdRushOrders} and \textbf{AvgRushOrders}: Moving standard deviation and average of the volume of rush orders in each chunk of the moving window.
  \item \textbf{StdTrades}: Moving standard deviation of the number of trades. 
  \item \textbf{StdVolumes} and \textbf{AvgVolumes}: Moving standard deviation and average of the volume of trades in each chunk of the moving window.
  \item \textbf{StdPrice} and \textbf{AvgPrice}: Moving standard deviation and average of the closing price.
  \item \textbf{AvgPriceMax}:  Moving average of the maximum price in each chunk.
  \item \textbf{HourSin}, \textbf{HourCos}, \textbf{MinuteCos}, \textbf{MinuteSin}: The hour and minute of the first transaction in each chunk. We encoded this feature with the sine and cosine functions to express their cyclical nature.
  \end{itemize}

Once a pump is detected, we pause our classifier for 30 minutes to avoid multiple alerts for the same event.

\subsection{The importance of rush orders}
In this section, we explore how rush orders are important to detect the start of a pump and dump operation.

\begin{figure} 
\includegraphics[width=1\textwidth]{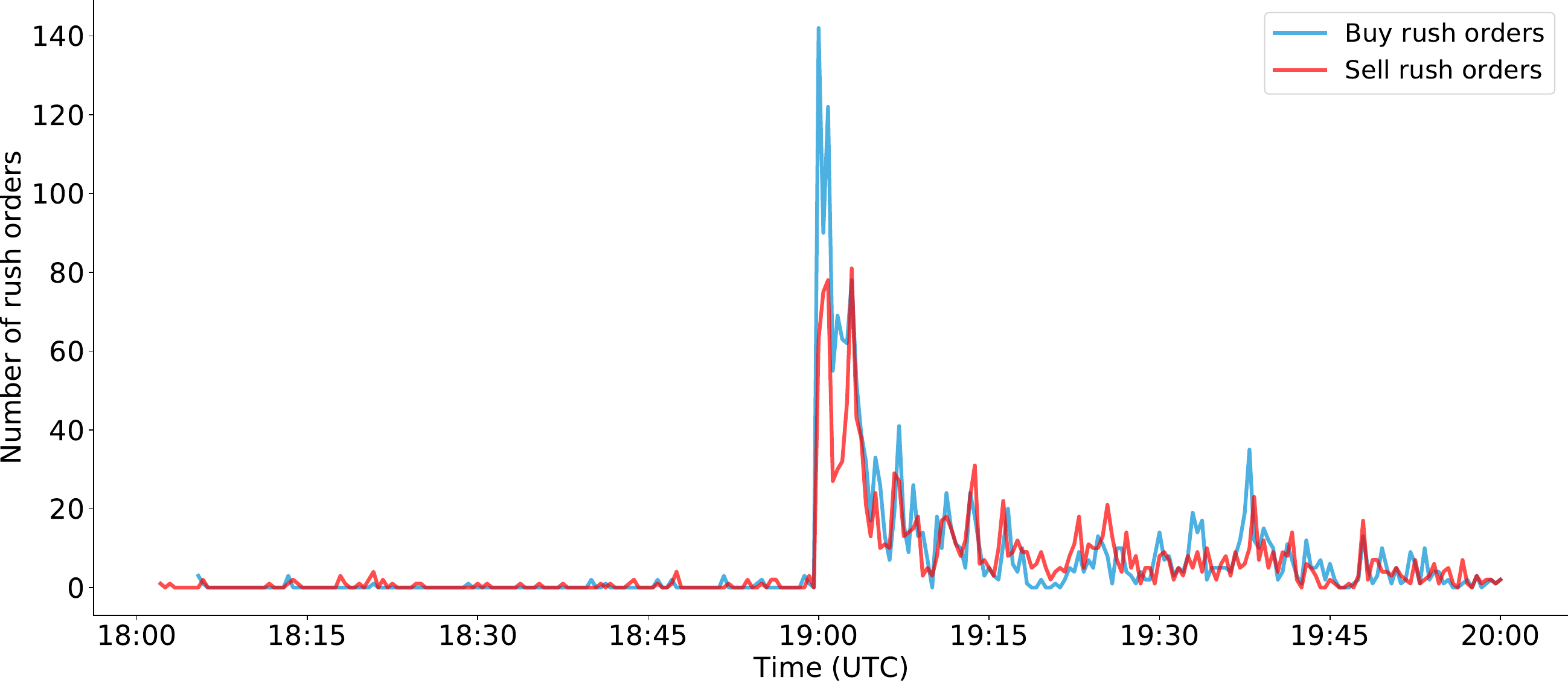}
\caption{Number of rush orders during the pump and dump on VIBE cryptocurrency.}
\label{fig:market_op}
\end{figure}

Fig.~\ref{fig:market_op} shows the number of buy and sell rush orders during a pump and dump scheme on the VIBE cryptocurrency on September 9th, 2018. 
As we can see, rush orders are rare during the hours before the pump and suddenly grow just at the start of the scheme. 
Comparing the number of buy and sell rush orders, we notice that buy rush orders are more prevalent than sell rush orders at the start of the pump operation. This is expected since the first part of the operation, the pump phase, consists of buying the asset as quickly as possible.
For this reason, we consider only the number of buy rush orders as a feature for our machine learning models.
Moreover, sell rush orders may indicate other phenomena (e.g., panic selling) and lead to false positives.

\begin{figure} 
\includegraphics[width=1\textwidth]{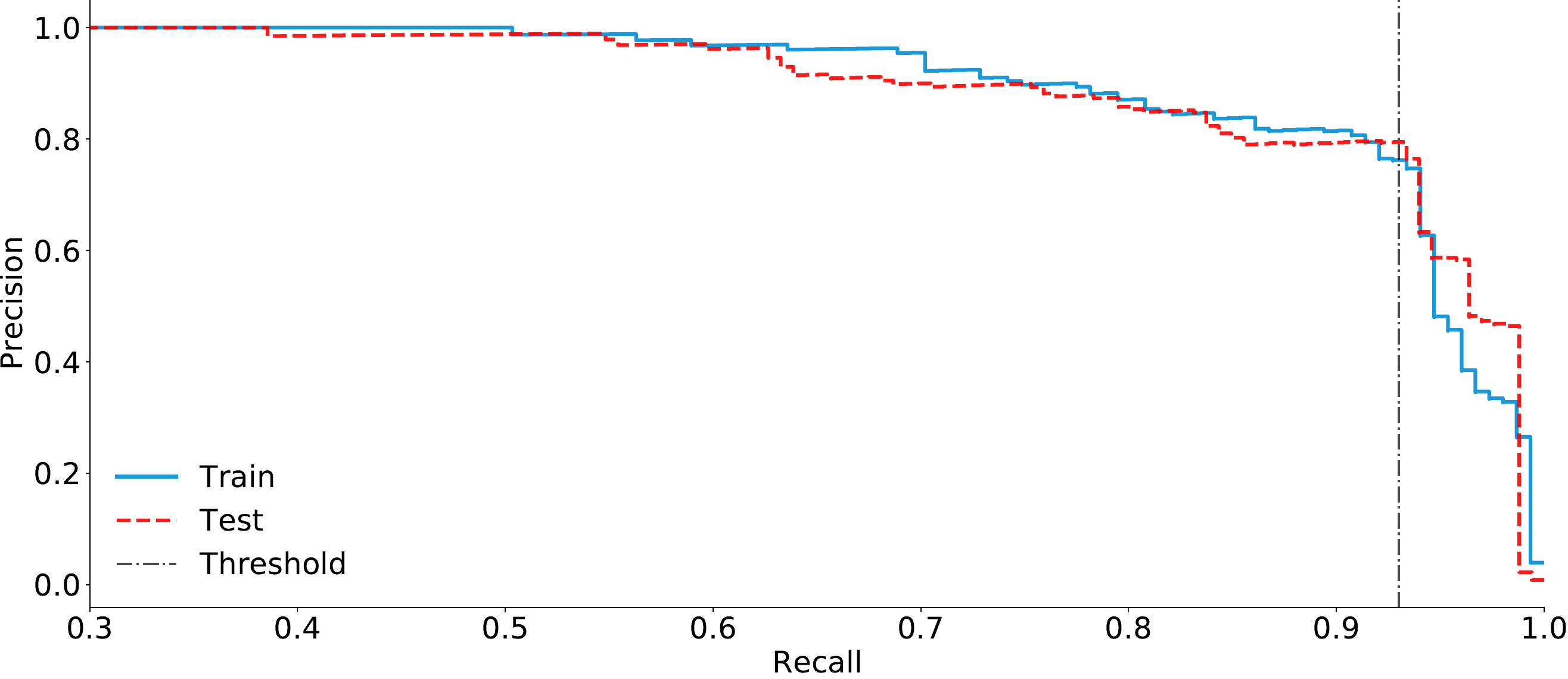}
\caption{Precision recall curve for train and test sets.}
\label{fig:pr_curve_train}
\end{figure}

We perform an experiment to understand if the rush orders are a practical feature to detect the beginning of a pump and dump scheme and find a threshold beyond which the growth can be considered abnormal. 
To learn the threshold, we proceed as follows: we compute the \textit{StdRushOrder} feature as described in Section~\ref{sec:features}. Then we label each chunk as True if the timestamp of the pump and dump signal falls into the chunk time range, False otherwise. We randomly split our dataset into the train (50\%) and test (50\%) sets, we compute the precision-recall curve for the train set, and we pick a threshold that is a trade-off between the precision and the recall. Then we evaluate the same metrics at the picked threshold for the test set.
Fig.~\ref{fig:pr_curve_train} shows the results. We choose 12.8 as the value for the threshold (the black dashed line in the figure). This value provides a precision of 81.2\% and a recall of 91.1\% on the train set (the blue line). As we can see, the same threshold also provides a very similar score on the test set (the red dashed line).
Given these results, we can claim that the rush orders feature is an excellent parameter to evaluate the start of a pump and dump.

\subsection{The results}

Although we retrieved 2 weeks of data for each pump and dump scheme, initially, we use only 3 days---the day of the fraud, the day before, and the day after. We can reasonably assume that no other scams are present for the same coin in this time frame. Indeed, among the market manipulations we collected, different groups arranged schemes on the same alt-coin a few days apart. However, we are aware that some groups delete the pump and dump signal from the chat history and that there are groups that we cannot monitor, such as groups that communicate in Chinese or Russian. Since our dataset consists of 317 pump and dumps, we do not split the dataset into the standard train test sets. We performed a 5 folds cross-validation to get a more reliable performance evaluation.
\begin{table}
\parbox{.50\linewidth}{
\caption{Classifiers performance with K-Fold cross validation.}

\centering
\begin{tabular}{l c c r c c c}

\toprule
Classifier & Chunk size & Folds & Precision & Recall &F1\\
\midrule
Kamps (Initial)& $1$ Hour & - & $15.9$\% & $95.3$\% & $27.2$\%\\
Kamps (Balanced)& $1$ Hour & - & $38.9$\% & $93.2$\% & $54.9$\%\\
Kamps (Strict)& $1$ Hour & - & $52.1$\% & $78.8$\% & $62.7$\%\\
\midrule
Random Forest & $5$ Sec & 5 & $94.6\%$ & $72.9\%$ & $82.4$\%\\
Random Forest & $15$ Sec & 5 & $96.4\%$ & $84.9\%$ & $90.0$\%\\
Random Forest & $25$ Sec & 5 & $98.2\%$ & $91.2\%$ & $94.5$\%\\
\midrule
AdaBoost & $5$ Sec  & 5 &  $90.0\%$ & $79.2$\% & $84.2$\%\\
AdaBoost & $15$ Sec & 5 &  $91.7\%$ & $87.7$\% & $89.7$\%\\
AdaBoost & $25$ Sec & 5 &  $95.4\%$ & $90.9$\% & $93.1$\%\\

\bottomrule
\end{tabular}

\label{tab:classifiers}
}
\hfill
\parbox{.45\linewidth}{
\centering
\caption{Features importance.}
\begin{tabular}{l r }

\toprule
Feature & Importance\\
\midrule
StdRushOrders  & $0.251$ \\
AvgRushOrders  & $0.123$ \\
AvgVolumes  & $0.081$\\
StdTrades  & $0.073$\\
StdVolumes   & $0.073$\\
AvgPriceMax    & $0.055$\\
AvgPrice    & $0.032$ \\
MinuteCos & $0.031$\\
MinuteSin & $0.022$\\
StdPrice    & $0.013$ \\
HourSin & $0.011$\\
HourCos & $0.003$\\
\bottomrule
\end{tabular}

\label{tab:importance}
}
\end{table}

For the Random Forest classifier, we use a forest of 200 trees and a maximum depth of 5 for each tree. 
Table~\ref{tab:classifiers} shows that the Random Forest classifier has outstanding results in terms of precision. However, the recall drops quickly, from 91.2\% to 72.9\%, when we reduce the chunk size from 25s to 5s.
To address this issue, we introduce a new approach with respect to the one used in previous work~\cite{la2020pump} that leverages an AdaBoost classifier. This approach is more balanced in terms of precision and recall and has better results in terms of F1-score.
Moreover, from the classifiers' results, it is possible to note the relationship between the chunk size and the performance of the classifiers. Indeed, while the precision is relatively stable in all the time frames, the recall increases as we increase the chunk size.

In Tab~\ref{tab:importance}, we list the importance, computed with the Gini Impurity, of each feature used with the Random Forest classifier. As we can see, the best are the ones based on the rush orders and the number of trades.
Once we defined our methodology, we trained a 25-second detector classifier with the $3$ day dataset and used the remaining two weeks of data (more than 14 millions 25-second chunks) as a test looking for other suspect events. After the evaluation, we got $86$ events that we are not able to link to evidence. 
Looking at the dynamics of the events, we believe that virtually all of them are pump and dumps whose evidence has been deleted or organized by groups that may not be public or that we cannot monitor.

\begin{figure}
\includegraphics[width=1\textwidth]{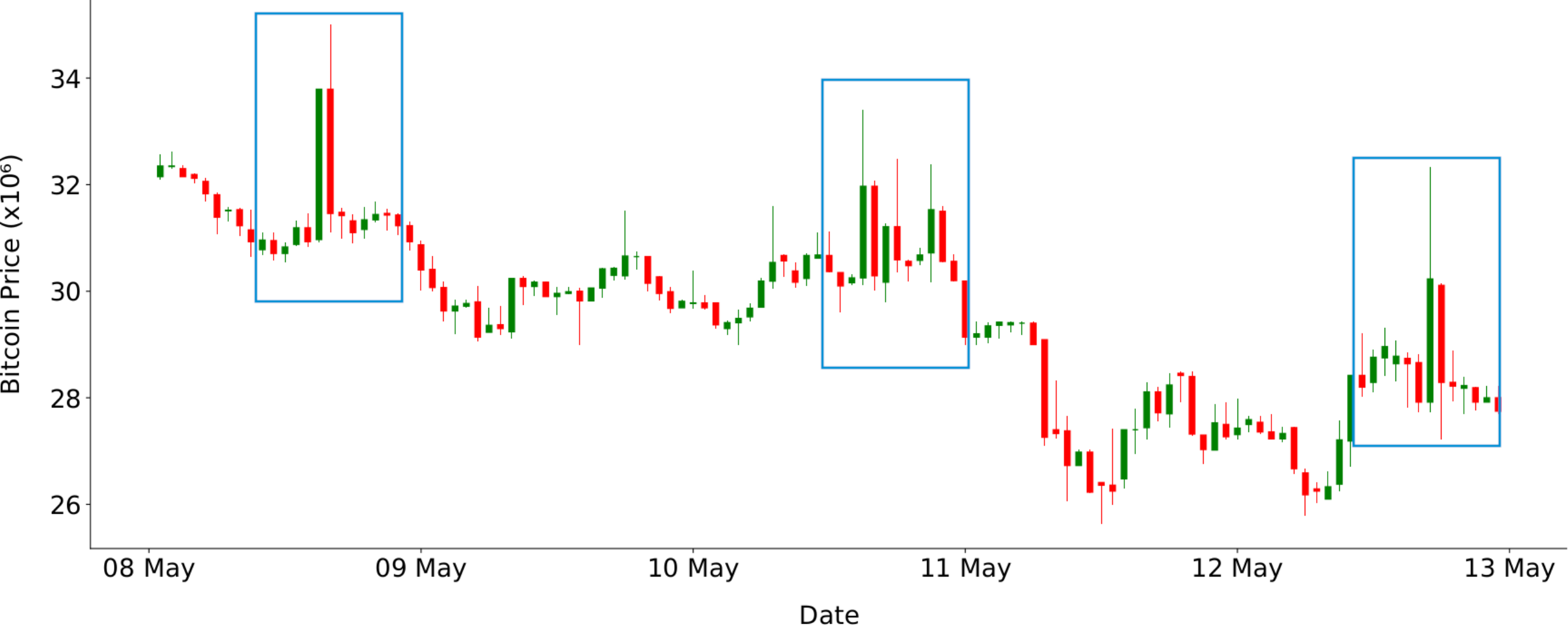}
  \caption{DLT candlestick chart.}
  \label{fig:false_positive}
\end{figure}

Fig.~\ref{fig:false_positive}, for example, shows the candlestick chart for the Agrello coin (DLT) from May 8 to 13. The event in the center is a pump and dump for which we have evidence. The other two are suspects detected by the algorithm. As you can see, the behavior is almost the same, including the fact that the currency quickly returns to its usual price (the dump). 
In any case, our classifier, based on the detection of the abnormal presence of rush orders and not just on the price, does a good job in detecting pump and dumps and suspect events that, anyways, the mindful investor wants to stay away from. 

\subsubsection{Long range experiment}

In the previous section, we found 86 alleged pump and dumps events we are not able to link to evidence. These events can raise some concerns about the use of our model in a real scenario.
Thus, we perform an experiment to assess the reliability of our detector over long time-frames.
We test our detector over three very different cryptocurrencies: Ethereum~\cite{buterin2014next}, Algorand~\cite{chen2019algorand} and Bread~\cite{bread}.
Ethereum and Algorand are, respectively, high and medium market-cap cryptocurrencies. As mentioned in Section~\ref{sec:discussion}, these assets are unlikely to be the target of pump and dump events. 
Thus, we can assume that every alert of our detector on these cryptocurrencies is a false positive.
Instead, Bread is a low market cap cryptocurrency with higher volatility. This means that this asset is more prone to quick market oscillations as well as market manipulations.
Moreover, it is the most targeted by pump and dump according to our dataset.
We consider all transactions performed on the three cryptocurrencies from their listing on Binance (2017-07-14 for Ethereum, 2019-06-22 for Algorand, and 2017-12-2 for Bread) to the end of the analysis (2021-01-31).
For Ethereum, our classifier finds 24 suspicious events over a period of 1,276 days. We obtain similar results on Algorand, where our classifier raises only 19 alerts on 591 trading days (on average, one false positive every month).
Finally, for Bread, we find 41 pump and dump events on more than 3 years of data, 24 of which are present in our dataset and 17 are suspicious. Thus, one suspicious event every 2 months.
Table~\ref{tab:long_range} summarize the results of our experiments.
At the light of this experiment, we believe that our detector can be handy in an real usage scenario, even if raises some false positive (less than 1 per month per monitored crypto).

\begin{table}
\small
\centering
\caption{%
Results for the long range experiment.
}\label{tab:long_range}
\begin{tabular}{l r r r r r }

\toprule
Cryptocurrency & Days analyzed & Events found \\
\midrule
Ethereum (ETH)  & 1,276 &  24 \\
Algorand (ALGO) & 591 & 19 \\
Bread (BRD) & 1,156 & 17\\
\bottomrule
\end{tabular}
\end{table}

\subsection{Comparison with other pump and dump detectors}
\label{subsec:comparision}
To the best of our knowledge, the best detector of pump and dumps in the literature is Kamps et al.~\cite{kamps2018moon}. We use their algorithm as the baseline. Their detector takes as input candlesticks of 1 hour. So, the best performance, in expectation, is of 30 minutes. Their methodology leverage two anomaly thresholds: Transaction volume and coin price. They compute the values of the thresholds using windows on the recent history of the candlestick under observation. If both the price and the volume are higher than the calculated thresholds, they mark the event as a pump and dump. Kamps et al. provide three different parameter configurations to compute the threshold: Initial, Balanced, and Strict. The Basic configuration maximizes the recall, the Strict the precision, while the Balanced is a trade-off of the two. In their work, they mention the number of alleged pump and dumps that their classifier detects. Unfortunately, they do not provide scores in recall and precision since their dataset lacks ground truth.

To use the Kamps et al. detector as the baseline for our task, we implemented their classifier and tested it on their dataset. We detected the same number of pump and dumps they report in their work. Then, we apply their methodology to our dataset---Table~\ref{tab:classifiers} shows the results. As we can see, all our classifiers outperform in terms of F1-score the Kamps et al. detectors. Our detector's performance is considerably better, we score 98.2\% of precision and 91.2\% recall against their 52.1\% precision and 78.8\% recall, and our detector is faster as well.
These results also show that, due to the cryptocurrency market's high volatility, the detectors based only on the coin price and transaction volume are prone to many false positives.

Differently from our work, Xu et al.~\cite{xu2018anatomy} build a classifier able to predict the next pump and dump's currency target to provide a tool for strategic trade. Since the goals are different, we can not make a comparison in terms of performance between our work and their solutions.
Indeed, they prefer to maximize the probability of gain from the investment, maximizing the recall at the expense of low precision. Xu et al. assume that buying wrong currencies does not affect the trading strategy because, on average, this does generate an economic loss.
Instead, in our case, we want to provide a reliable approach---with high precision and recall--- to help investors stay out of the market when a pump and dump scheme is in action and to analyze anomalies in historical data.

\section{The Crowd Pump}
Now, we focus on a new kind of pump operation. We will call it \textit{crowd pump}---a pump and dump event that results from the non-directly organized actions of a crowd of people. We analyze how these operations happen, and we illustrate the differences from standard pump and dumps. Lastly, we offer that it is possible to leverage our dataset to build a classifier that can also detect crowd pump events.

\subsection{A description of the crowd pump phenomenon}
In January 2021, the stock market was puzzled by an unprecedented rally of GameStop (GME). The GME stock had been gradually losing value for a couple of years, as sales of physical copies of video games plummeted due to the shift towards digital purchases~\cite{gamestopdrop}. During the COVID-2019 pandemic, the situation worsened to the point that GameStop announced it would close more than 1,000 stores by April 2021~\cite{gamestopclosestores}.
GME quickly became easy prey for short-sellers, economic agents that bet on the fall of specific securities. Short-sellers borrow stocks, sell them, and buy them later, when the price is expected to be lower, to give them back to the lender. 

This operation would have gone unnoticed, as this market practice, albeit somewhat controversial, is common. The turning point came when a group of users active on Reddit~\cite{reddit}, one of the most popular social news aggregation and discussion websites~\cite{redditpopular}, started to buy large quantities of GME stocks.
These users communicated in a \textit{subreddit}---a user-created board that covers a specific topic---called \textit{r\textbackslash{wallstreetbets}}.
Initially, the users started to invest in the GME stocks because they believed they were undervalued. Only later they began to do it as a political statement against hedge funds~\cite{sendmessage}.
The operation was a great success and the subreddit users managed to raise the stock price of GME by more than 1,900\%, from \$17.25 on January 4 to \$347.51 on January 28~\cite{redditGME}. Due to the media interest, the subreddit gained more than 3 million followers in that period. GME became the most traded stock in the U.S. stock market on January 26~\cite{gmetoppingstocks}.
Due to the results of this operation, people started collaborating to buy other stocks such as AMC (AMC Entertainment Holdings), BB (BlackBerry Ltd.), and NIO (NIO Inc.). Their prices increase rapidly in a few days~\cite{otherstocks}. In response, several digital trading services like Robinhood began restricting trades on the stocks that were getting pumped~\cite{robinhood}.

Due to these limitations, the attention moved to cryptocurrencies---less regulated and still with a combined market capitalization that topped \$1 trillion~\cite{cryptorise}.
The first coin to get widespread attention was Dogecoin. Dogecoin was originally founded as a joke on December 6, 2013~\cite{whatisdogecoin}. The price of the coin skyrocketed on January 28, 2021, after a Reddit group, called \textit{\textbackslash{SatoshiStreetBets}}, proposed to make it the equivalent of GME for the cryptocurrency market. Dogecoin had an increase in the price of over 800\% in 24 hours, from \$0.0077 to \$0.07 according to data of CoinGecko~\cite{coingeckodoge}.
The price increased in several distinct phases, driven by the tweets of well-known personalities like Elon Musk, rock star Gene Simmons, and rapper Snoop Dogg, reaching its highest value ever of \$0.079~\cite{dogetweets}.

The second target was the Ripple (XRP) crypto-coin.
At the time of these events, the XRP suffered a challenging moment due to a lawsuit that started on December 22, 2020. The SEC accused Ripple of performing illegal security offerings of \$1.3 billion in XRP for seven years beginning in 2013~\cite{xrpsecproblems}.
This action caused a drop in the coin price from \$0.42 on December 22 to \$0.18 on January 4. Several exchanges delisted XRP. Including Coinbase, one of the largest~\cite{xrpdelisted}.
The delisting reduced the liquidity of the coin significantly, creating the perfect breeding ground for market manipulations~\cite{xrplowliquidity}.
In this case, the operation was organized on a Telegram group called "Buy \& Hold XRP FEB 1st, 2021" that was later renamed "BUY \& HOLD XRP FEB 1st, 2021 @8:30AM"~\cite{xrpteelegram}.
The group grew exponentially in the 24 hours following its creation, reaching the limit of 200,000 members of Telegram.
The group aimed to buy massive quantities of XRP at a precise date and hour---February 1, 2021, at 13:30 UTC.
However, many members started buying it massively the days before the pump, and the cryptocurrency jumped $56\%$ up in price, reaching the biggest single-day percentage gain since December 21, 2017~\cite{robinhood}. So, the price was already high at the pump, and the group could not increase it any further.

\subsection{Analysis of crowd pumps}
\label{sec:crowd_pump_analysis}
Although it is well-known that the DogeCoin pump starts from some popular subreddits~\cite{dogestreetbets}, it is unclear who started the pump and how they carried out the operation.
We analyze all the Reddit users' posts on the subreddits mentioned above to answer these questions.
A \textit{submission} is the first post of a new discussion thread and may contain links, text, and images.
To perform our analysis, we downloaded all the submissions from January 01, 2021 to February 02, 2021 of some popular crypto-related subreddits: \textit{r\textbackslash{SatoshiStreetBets}, r\textbackslash{WallStreetBets}, r\textbackslash{Cryptocurrencies}, r\textbackslash{DogeCoin}} subreddits. We downloaded data from these subreddits since in the period the terms \textit{"Doge"} and \textit{"DogeCoin"} appear mainly in them, according to the Redditsearch tool~\cite{redditsearch}.
To retrieve the submissions, we leveraged Pushshift~\cite{baumgartner2020pushshift}, a service that provides access to Reddit data overcoming the limit of 1,000 posts of the official APIs.

We globally retrieved 656,146 submissions, of these 626,700 (95.5\%) from \textit{r\textbackslash{WallStreetBets}}, 23,485 (3.6\%) from \textit{r\textbackslash{SatoshiStreetBets}}, 5,443 (0.8\%) from \textit{r\textbackslash{Cryptocurrencies}} and lastly 518 (0.1\%) from \textit{r\textbackslash{DogeCoin}}.
From the downloaded data we took into account only the submissions that contain the name of the coin ("DOGE", "DogeCoin") and some of their very popular variations used in the cryptocurrencies slang, such as: "DOGIE", and "DOGUE". In the end, we get 27,868 submissions with the following partition: 19,016 (68.2\%) from \textit{r\textbackslash{WallStreetBets}}, 8,383 (30.1\%) from \textit{r\textbackslash{SatoshiStreetBets}}, 194 (0.7\%) from \textit{r\textbackslash{Cryptocurrencies}}, and 275 (1\%) from \textit{r\textbackslash{DogeCoin}}. Finally, we study the message distribution over time and their relationship with the price of DogeCoin.

\begin{figure}[h]
\includegraphics[width=1\textwidth]{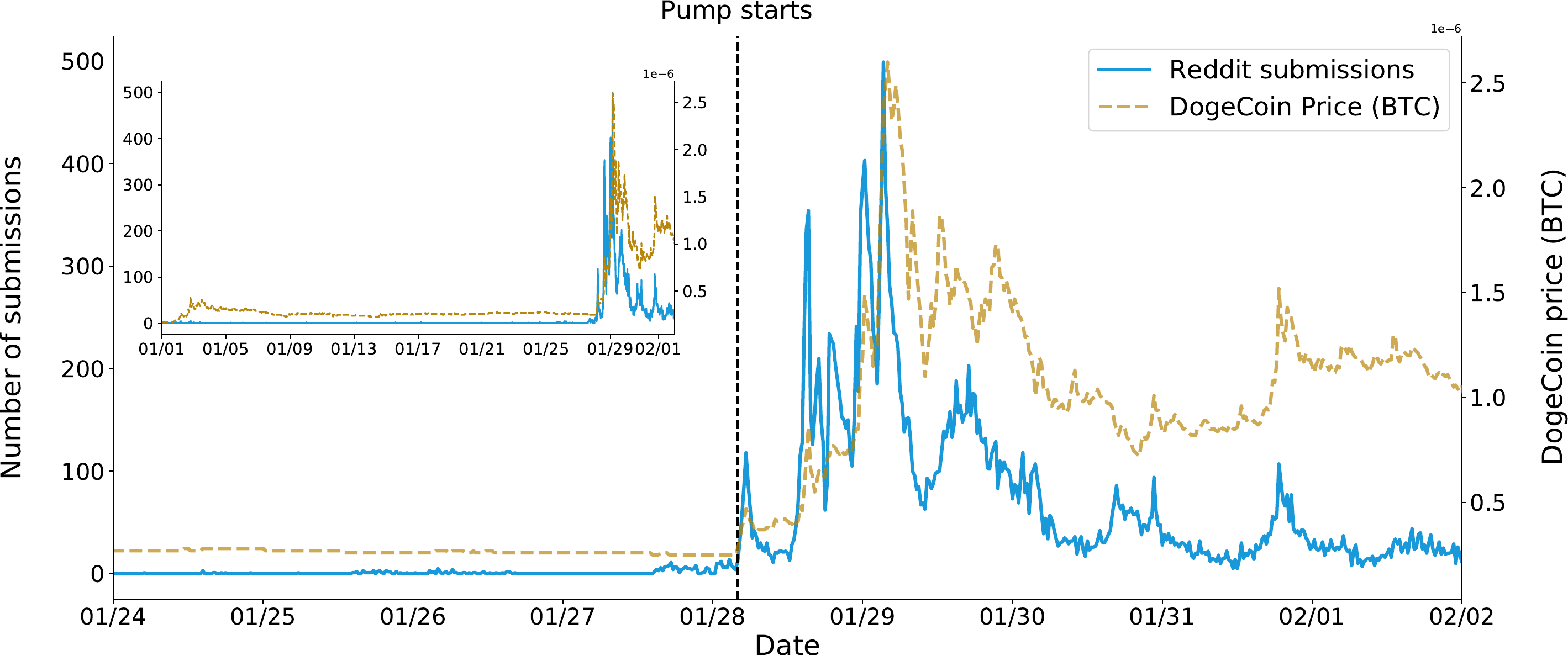}
\caption{Number of submissions mentioning DogeCoin sent on the subreddits vs. DogeCoin price in BTC.}
\label{fig:doge}
\end{figure}

Figure~\ref{fig:doge} shows the number of submissions posted in the subreddits that mention DogeCoin (solid blue line) and the price of DogeCoin in Bitcoin (dashed gold line). As we can see in the upper left chart of the figure, subreddits rarely mention the coin in the weeks before the pump, and the price is stable. In the 24 hours before the pump (vertical dashed line), it is possible to note that some submissions about the coin begin to pop up steadily. However, the price is still stable. After the vertical dashed line, the coin gets a massive spike in popularity, and the price abruptly rises. From this moment, the price of DogeCoin and the number of submissions on Reddit follow the same pattern.

In the light of this analysis, we dig into the posts before the pump. The goal is to understand how the users arranged the operation.
We find out that most of these posts tried to drum up the attention on the DogeCoin proposing to pump the currency. Initially, the users did not welcome these posts. The administrators often removed the content because it violated the netiquette of the subreddit. Among these submissions, we found a particularly interesting one on the r\textbackslash{DogeCoin} subreddit. Here, a few users were trying to arrange a pump on the DogeCoin on January 28 at 10 AM, $5$ hours later than the actual start of the pump. 
Nonetheless, none of these submissions had any effect on the price of the DogeCoin, as shown in Fig.~\ref{fig:doge}.
In our opinion and news~\cite{dogestartcnbc}, the message that triggered the rally of the DogeCoin, for timing and users welcoming, was posted on January 28, 2021, at 4:05:50 UTC and states: \textit{"Let's make DOGIECOIN a thing. That's it, that's the post"}. The submission had only the title, no message body, and no picture.

To better understand why this message triggered the pump, we investigated the creator of the submission, expecting her to be popular on the Reddit community. Surprisingly, we found out that, although the user is very active on Reddit with more than 854 submissions and 769 comments, only 4 submissions (0.4\%) and 17 comments (1\%) are related to crypto or finance. Thus it is doubtful that the author is a crypto-influencer, and it is hard to understand why so many users followed this message.

We performed a similar analysis also on the crowd pump carried out on the Ripple cryptocurrency.
For this case study, we analyze the messages on Reddit in the same time frame we did for the DogeCoin, since the two events occurred within a few days of each other.
We consider the same subreddits of the previous analysis, with the exception of \textit{r\textbackslash{DogeCoin}} subreddit and including the \textit{r\textbackslash{XRP}} (5,444 submissions), obtaining globally 661,072 submissions.

\begin{figure}[h]
\includegraphics[width=1\textwidth]{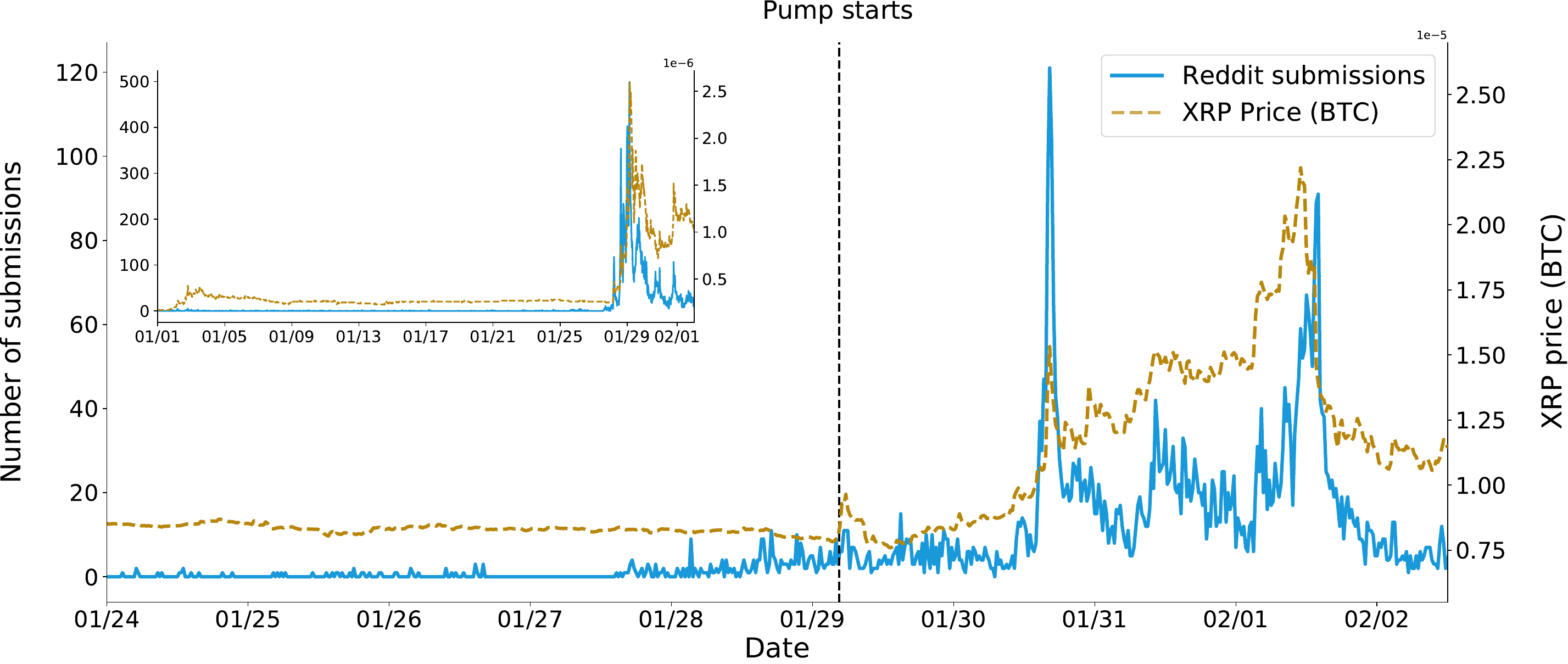}
\caption{Number of posts mentioning XRP vs XRP price in BTC.}
\label{fig:xrp}
\end{figure}

In this case, we focus on the submissions that mention one of the cryptocurrencies.
Figure~\ref{fig:xrp} shows the number of posts in the subreddit that mention Ripple (solid blue line) and the price of Ripple (dashed gold line). As we can see, the coin is rarely mentioned in the weeks before the pump, while it starts to get attention in the days before the pump. Similar to what happened in the case of the DogeCoin pump.
Reading these messages, we find out that the cause of this increase in the posts is due to Redditors driven by anti-SEC sentiment and inspired by the DodgeCoin and GME pump operations.
The birth of the Telegram group \textit{"OFFICIAL BUY \& HOLD XRP"} gathered these users, and group members began to promote the group itself.
Different from the DogeCoin crowd pump, where the number of posts on Reddit and the cryptocurrency price seem to follow the same trend, in this case, the two lines seem to be more independent, except for the price peaks.
Analyzing the beginning of the pump (solid dashed line in Fig.~\ref{fig:xrp}), it is possible to note that the price quickly rises while the number of submissions on Reddit does not. Some hours later, the price returns to its real value (January 29 at 5:00 UTC), and then the price increases again (January 30 at 16:00 UTC).

This behavior makes us suspect that the pump does not start from Reddit. Thus, we investigate the messages sent on the Telegram group, for which we were able to export all the messages, files, videos, and images. Since, to the best of our knowledge, the group is no longer accessible, and the group chat is not publicly available, we publicly release it as a further contribution~\cite{xrpchat}. 
The Telegram group counted exactly 200,000 members and 45,548 messages. We do not know when the group was created, but the first message appeared on January 28 at 20:19:09 UTC.
Unlike pump and dumps, the organizers did it on a Telegram Group instead of a Telegram Channel. Hence, all the group members could write in the chat, not only the admins. After the creation of the group, the chat was open, and the members could freely talk about the event and how to participate.
However, the situation escalated around January 29 at 5:00 UTC. From this moment, maybe for a slight fluctuation of the Ripple's price or an extra-group coordinated action of a set of users, the members start to urge the chat to \textit{BUY!} the coin, starting the pump way earlier than expected.
This event occurs almost at the same time as the first spike in price that we see in Figure~\ref{fig:xrp}.
The admins promptly reacted by turning off the chat and resumed it only twice before the day of the pump---the first time on January 30 at 20:05 UTC, the second one on January 31 at 6:03 UTC.
In both cases, the chat opened only for 30 minutes, and the admins asked the member of the groups to indicate from which countries they were posting. Then, the chat was opened again 9 hours before the pump for a few seconds. As discussed before, the pump was a failure as the group could not further raise the price of the coin.

At the end of our analysis, we find the following main differences between crowd pump and pump and dump operations:
\begin{itemize}
    \item \textbf{Different goal:} The aim of a crowd pump is not to inflate the price of an asset and sell it to scam unaware investors. In these kinds of operations, the organizer and part of the community often encourage the participants to hold their stock to keep the value high. We noticed this attitude in both the crowd pump events carried out on the crypto market.
    A clear example is the crowd pump organized on the XRP currency.
     In this case, the group creator clearly states in the Telegram group chat that the operation aims to hold the currency. The admin also publishes a disclaimer video on his Youtube channel explaining the purpose of the group. Quoting the description of the video: "This is not a "pump and dump" group. This is a community-led event to bring awareness to the XRP ledger"~\cite{xrpvideo}.
    
    \item \textbf{Lack of coordination and leadership:} 
    Even if we saw on both the crowd pump events attempts to coordinate to buy at a specific hour, they always failed.
    Unlike standard pump and dump, the organizers reveal the coin to pump in advance. Thus, people start to buy the coin in advance or when they believe the operation has begun. A simple fluctuation of the market or a single post can trigger a ripple effect that leads to the start of the pump.
    
        \item \textbf{Different time frame and price increase rate:} 
    As we saw, in standard pump and dump, the operation lasts for a few minutes or rarely for a few hours, and the price grows almost immediately.
    In a crowd pump, the price increases abnormally, but it takes hours or days before the coin reaches its maximum peak. This behavior is due to several factors. The goal is different, and some investors do not immediately sell the coin to take a profit. No one knows when the pump will start. Therefore it can take time before the crowd realizes that the operation has begun. Finally, the news and influencers work as an echo chamber, and more and more people join the process making the price of the coin increase in waves.
    Consequently, while in standard pump and dump the price of the coin returns to its natural level as the event ends, in crowd pump and dump, after more than a month\footnote{March 2, 2021}, the price of the DogeCoin is still 500\% higher than its pre-pump value, and the XRP is stll 100\% higher.
\end{itemize}

\subsection{Crowd pump detection}

\begin{figure}
\includegraphics[width=1\textwidth]{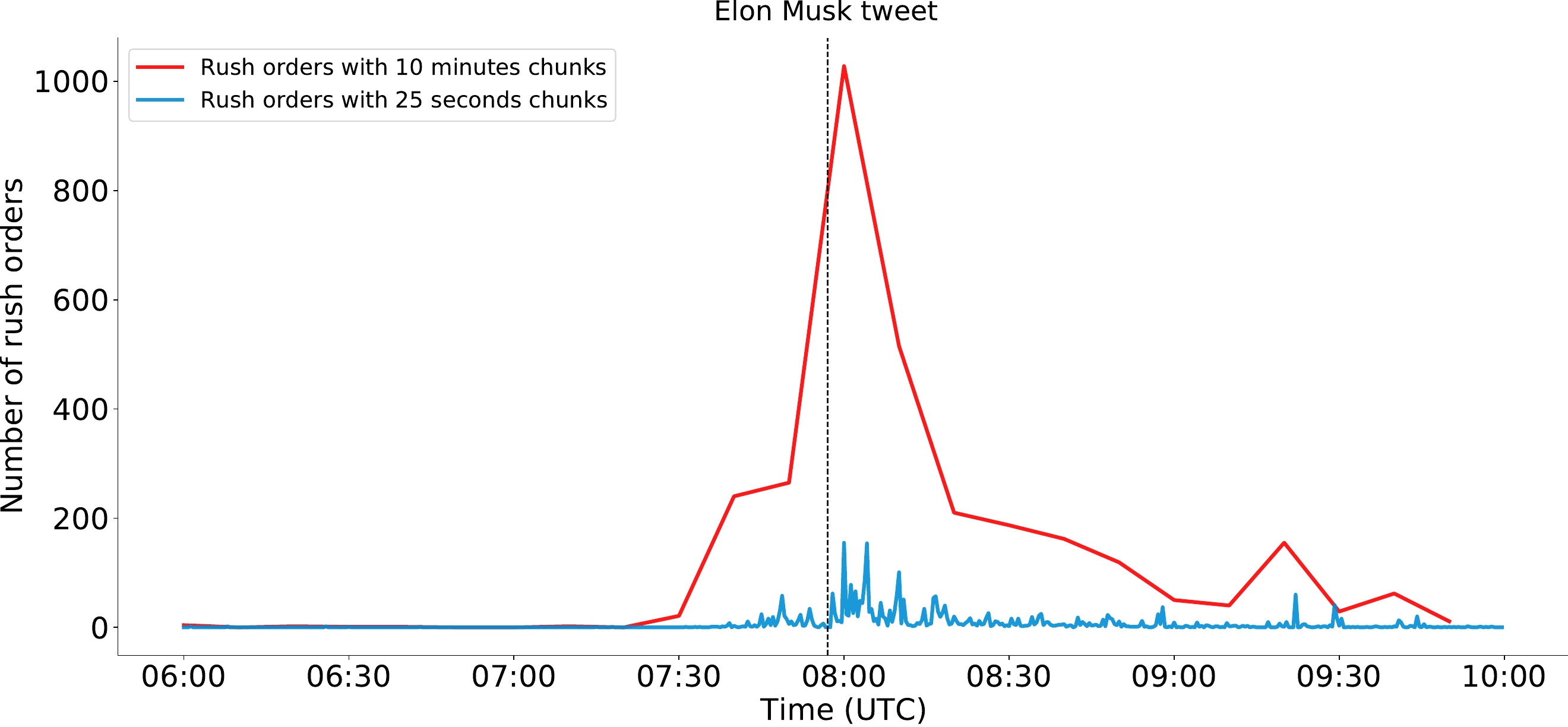}
\caption{Number of rush orders before and after the tweet of Elon Musk about Dogecoin (February 4 at 7:57 UTC).}
\label{fig:elontweet}
\end{figure}

In this section, we assess the potential of our machine learning model in detecting crowd pump operations.
Although there are some key differences between the crowd pump and standard pump and dump, our intuition is that the rush orders are a very relevant feature also in this kind of operation.

In particular, we consider the number of rush orders in an interval of two hours around the publication of a tweet of Elon Musk that shill the DogeCoin~\cite{dogetweets}.
We make this choice because, in this case, we have the timestamp of the tweet and we can be sure about the moment in which the operation starts.
Figure~\ref{fig:elontweet} shows the number of rush orders in two hours around the publication of the tweet. The purple line represents the number of rush orders grouped in chunks of 25 seconds, while the red line in chunks of 10 minutes. In the figure, it is possible to note a considerable amount of rush orders after the tweet, precisely like in pump and dump events after the admin announces the target coin.
However, looking at the purple line (25 seconds chunk), we find that the pattern of the rush orders is very different from the one we see for the standard pump and dumps (Figure~\ref{fig:market_op}). 
Indeed, there is no neat big spike in the number of rush orders, but a gradual increase with several small spikes.
This behavior is not surprising. There is no synchronization of the investors---they jump into the market in waves depending on when the message hits the social platforms on the web and when they see it.

Due to this different behavior, our detector trained on the standard pump and dumps cannot capture the crowd pump analyzing short chunks of transactions. Moreover, we cannot efficiently train a new detector for the crowd pump operations because of the lack of a dataset. However, expanding the chunk's time frame size makes it possible to collapse the different waves of rush orders into a unique chunk and get a well-outlined spike. The red line in Figure~\ref{fig:elontweet} shows the number of rush orders grouped in chunks of 10 minutes. Here, we can see that the pattern is very similar to a pump and dump operation, like the one we reported in Figure~\ref{fig:market_op}, and thus it is now reasonable to think that our detector can find these kinds of events.

\subsubsection{The new model}

To detect the crowd pumps, we trained a new classifier based on the Random Forest algorithm like the one used to detect standard pump and dumps. This time we trained the model on the full dataset (317 pump and dump events) described in Sections~\ref{sec:dataset}. We used the same feature we leveraged to build the previous detector, except for the one related to the time. We removed these features because they are specifically tailored for the standard pump and dumps carried out by Telegram groups. The new detector achieves an F1-score of 89.4\% in 5 fold cross-validation.
In the case of crowd pumps, we test our approach only on two events: XRP and DOGE.
For the training phase, we used 25 seconds chunks. 
Instead, we aggregate the trading data in chunks of 10 minutes for the test phase.
After detecting an event, we pause our classifier for 6 hours to avoid multiple alerts. In this case, we pause the classifier longer than we did for the standard pump and dumps because the operations last more time.

\subsubsection{The Dogecoin pump}

To find out if our detector can catch the start of the Dogecoin crowd pump, we downloaded all the transactions from Binance from January 1 to February 10, 2021.
Even though we know that the pump happened on January 28, 2021, we run the detector for some weeks before the pump to check if any suspicious activity is detected and to validate the classifier's robustness on false positives.
At the end of the execution, our classifier detects the following 5 events:
\begin{enumerate}
    \item  \textbf{January 2, 2021, at 3:00 UTC:} At first sight, the event seemed a false positive. However, after a search on the web, we found that the news~\cite{angelawhite} reported a price surge of the DogeCoin driven by a tweet from the adult film star Angela White. The actress stated that she is a DogeCoin investor since 2014. The tweet features a photo of the actress wearing a T-shirt with a Shiba Inu image, the DogeCoin mascot, and received more than 10,000 likes.

    \item \textbf{January 28, 2021, at 4:10 UTC:} 
This alert falls exactly in the same chunk of the Reddit post that sparkled the DogeCoin popularity in the \textit{r\textbackslash{SatoshiStreetBets}} subreddit, discussed in Section~\ref{sec:crowd_pump_analysis}.

\item \textbf{January 28, 2021 at 14:20 UTC:} It is not easy to link this warning to an individual event. However, investigating on Reddit and Twitter, we find on Reddit an abrupt increase of messages that mention the coin (see Figure~\ref{fig:doge}). Moreover, between 14:00 UTC and 15:00 UTC in the U.S.A., the hashtag \textit{"\#dogecoin"} became a trending topic on Twitter, with more than 91,000 tweets. 2 hours later \textit{"\#dogecoin"} became a worldwide trending topic, accordingly with the data provided by ExportData.io~\cite{exportdata} and TT-History~\cite{TT-History}. In our opinion, it is safe to assume that this alert detected many investors that have flooded the market.

\item \textbf{January 28, 2021, at 23:40 UTC:} This is very likely due to Elon Musk's tweet (January 28, 2021 at 22:47 UTC) on Dogecoin.
In particular, the tweet contains a picture that mimics the Vogue magazine with a dog picture on the cover, and the title of the magazine changed to "Dogue." More than 450,000 users liked this tweet.  
The detector raised the alert about one hour later. However, looking at the price evolution of the DogeCoin in the hour following the tweet, it is possible to note that investors enter into the market slowly. Indeed, at the time of the tweet, the price of the DogeCoin was at \$0.024; at the time of the alert, the price was \$0.03 (+25\%). The coin reached its first peak in price one hour later, touching \$0.05 (+108\%). Then, around the 4:00 UTC of January 29, the coin achieved \$0.08 (+233\%) its maximum price of the month.

\item \textbf{February 4 at 8:00 UTC:} This alert is also related to a Tweet of Elon Musk. This time he posted a tweet that contains a meme portraying him as Rafiki from the Lion King---the animated movie, standing on Pride Rock and raising a Doge-headed Simba~\cite{dogemusk}. In this case, our detector captured the abnormal market movements 13 minutes after the tweet has been posted (\textit{i.e.} the very first chunk computed after the tweet). Unlike the previous tweet, this one gets much more attention, with more than 1 million likes on the social network, and the market reacts faster. In this case, our classifier detects the event when the price of the DogeCoin was at $\$0.04$, while the coin reaches its price peak at \$0.06, almost one hour later. 
\end{enumerate}

\subsubsection{The Ripple pump}

Again, for the Ripple crowd pump, we run our classifier on all the transactions closed on the Binance exchange from January 1 to February 10, 2021. In the considered time frame, the detector raises the following 4 alerts:
\begin{enumerate}
    \item \textbf{6 January, 2021, at 14:40 UTC:} 
    To the best of our knowledge, this alert is not related to the Reddit community. Instead, the news that a petition to the White House to stop the SEC lawsuit against Ripple hits 35,000 signatures~\cite{xrppetition} has probably caused new trust in the XRP coin. The price went from \$0.23, at the moment of the alert, to \$0.37 (+38\%) of the following day.
\item \textbf{19 January, 2021, at 5:50 UTC:}
This is the exact moment when several exchanges, including Coinbase, the $3^{th}$ exchange by volume of transactions, delisted XRP from the trading pairs~\cite{xrpdelisted}. The delisting follows the SEC lawsuit. Two hours after the alert, we record an abrupt rise in transaction volume on Binance and the price from $\$0.29$ to $\$0.33$ ($+9\%$). The alert is probably due to trading bots or investors that moved their assets from one exchange to another.
\item \textbf{29 January, 2021 at 5:00 UTC:}
This is when we noticed the excitement in the "OFFICIAL BUY \& HOLD XRP" group, with the members of the group that start to urge to buy the coin. As we discussed in Section~\ref{sec:crowd_pump_analysis}, the users' excitement comes together with the beginning of the XRP rally.
\item \textbf{30 January, 2021 at 16:00 UTC:}
For this alert, we do not have clear evidence of what triggered the event. However, in the hour before this alert, the Ripple cryptocurrency starts to hit the news, becoming one of the most searched words worldwide on Google ~\cite{googletrendxrp,googletrendxrp2}. At the same time, the number of posts on Reddit about the Ripple cryptocurrency increased dramatically. Driven by the news, an odd number of investors may have joined the market and started to buy in a rush the currency to avoid missing a good profit opportunity, triggering our detector.
\end{enumerate}

It is important to note that when the pump was scheduled (February the 1st, 2021, at 13:30 UTC), the detector did not raise any alert. This is not surprising since, as discussed before, the pump failed~\cite{xrpfailed}. 
Thus, the classifier detected the start of the pump two days before, catching the users that bought the coin in advance.

Looking at the results we achieved, we believe that our first attempt to build a classifier to detect crowd pump events shows excellent results. Nonetheless, we could further improve our detector by combining features from social media and related to the market exchanges' financial transactions.

\section{Related work}
The pump and dump phenomenon is older than the cryptocurrency revolution. Therefore, a vast portion of the literature is about pump and dumps done in the traditional stock market. Allen et al.~\cite{allen1992stock} identify three categories of market manipulation schemes: Information-based, action-based, and trade-based. The pump and dump schemes are usually a combination of information-based and trade-based manipulation.
In 2004, Mei et al.~\cite{mei2004behavior} show that it is possible to carry out pump and dump schemes just leveraging the investors' behavioral biases. They test their theory on the pump and dump cases prosecuted by the SEC from 1980 to 2002, which confirms their hypothesis.

Several case studies highlighted that emerging markets were prone to pump and dump schemes. Khwaja et al. in~\cite{khwaja2005unchecked} show that the limited Pakistani regulation on the national stock exchange allowed brokers of the Karachi Stock Exchange to arrange pump and dump schemes.
Jiang et al.~\cite{jiang2005market} investigate the stock pools scheme of the '20s using daily trading volume from the New York Stock Exchange between 1927 and 1929. The stock pools are groups of traders that delegate to a single manager to trade stocks on their behalf. Since a pool can move a large amount of money, it can increase the volume of trades and attract outsiders to the market. When the stock pool exits the market, the price quickly drops. 
As reported by the University of Innsbruck in~\cite{frieder2007spam}, the internet boom in the early years of 2000 led to the birth of a new email-based pump and dump scheme. In this new kind of fraud, the manipulators secure their position on the market and then send millions of e-mails claiming to have private information about substantial increases in the prices of specific stocks. After luring new investors, and the price higher, the manipulators sell the security and stop the spam campaign.
A subsequent analysis in 2013 by Siering in~\cite{siering2013all} shows that despite the authorities have taken several countermeasures against fraudulent stock recommendations, email-based pump and dump campaigns are still flourishing. 

The work of Gandal et al.~\cite{gandal2018price} show evidence that the first price spike to \$1000 of the Bitcoin may be market manipulation. Using the well-known dataset of the Mt.Gox exchange, they found suspicious trading activities carried out by two actors, named 'Willy bot' and 'Markus bot.' The purpose of these actors was to buy Bitcoin to increase the price and the daily volume artificially.
Krafft et al.~\cite{krafft2018experimental} investigate the behavioral patterns of the users on the Cryptsy exchange market. In their work, they show that even tiny volumes of buy trades can influence the market. They use bots to buy a small number of random currencies and conclude that traders tend to buy coins with recent activities.
Li et al.~\cite{li2018cryptocurrency} conduct an empirical investigation on trading data obtained from the pump and dumps from Binance, Bittrex, and Yobit, focusing on the economic point of view. They show that pump and dumps lead to a short-term increase in prices, volume, and volatility followed by a reversal of the trend after some minutes. Moreover, they show that the investors' gain depends critically on the time they obtain the signal. For this reason, outside investors are systematically disadvantaged.
Victor et al.~\cite{victor2019cryptocurrency} perform quantification and detection of pump and dump schemes coordinated through Telegram and executed on Binance. They test their machine learning model considering 125 pump and dump collected from Telegram as confirmed pumps and the 20 most retweeted tweets of the official Twitter accounts belonging to each coin as negative samples.
They do not aim to catch a pump and dump in real-time as their feature considers a 30 minutes interval and tries to capture both the pump and the dump phase.
Hamrick et al.~\cite{hamrick2018economics} conduct an analysis on Discord and Telegram, identifying more than 5,000 pump and dumps from January 2018 to early July 2018. However, they use a different definition of pump and dump, including events that we define 'signals.' With our definition, they found 704 pump and dumps. They measure the factors that lead to the success of a pump, defined as the increase in the price of the coin. Some of the most important are the volatility of the coin and the number of people in the groups.
Dhawan et al.~\cite{dhawan2020new} study 355 cases of pump and dumps in the cryptocurrency markets. They show that pumps generate an average price distortion of 65\%, abnormal trading volumes in the millions of dollars, and enormous wealth transfers between participants. They highlight that this kind of manipulation is likely to persist as long as regulators and exchanges turn a blind eye.
Nizzoli et al.~\cite{nizzoli2020charting} conduct a study on 50M messages collected on Twitter, Telegram, and Discord. They highlight the existence of two different manipulations: Pump and dump and Ponzi schemes. They found that 56\% of crypto-related Telegram channels are involved in manipulations and that bots massively broadcast these deceptive activities.
Chen et al.~\cite{chen2019detecting} develop an apriori algorithm to detect pump and dump on Bitcoin using the leaked transaction history of Mt. Gox Bitcoin exchange. They do not have a ground truth of confirmed pumps. Thus, they try to find groups of users who usually buy or sell the asset simultaneously. This is possible thanks to each user's complete transaction history--information typically unavailable and protected by privacy.
The work of Kamps et al.~\cite{kamps2018moon} shows a first attempt to detect pump and dumps using an adaptive threshold. They bring up the issue that a reliable dataset of the confirmed pump and dumps scheme does not exist, so they can not fully validate their results. A contribution of our work is to release such a dataset. Xu et al.~\cite{xu2018anatomy} focuses on the difficult task of predicting pump and dumps, using one-hour intervals data from Cryptopia and Yobit, also showing an approach to leverage the prediction to invest in alt-coins. Since both works have some goals in common with ours, we conducted a thorough analysis of their results on subsection~\ref{subsec:comparision}.

\section{Discussion}
\label{sec:discussion}
\noindent\textbf{Is it possible for pump and dump groups to avoid detection?} We based our features on the abnormal change of some market parameters and, at the same time, to be robust against the natural oscillations of the volatile cryptocurrency market. If the admins of groups or other members buy the currency gradually, and the users are few, our classifier may not detect the pump and dump. Indeed, our classifier cannot detect four of the pump and dumps in our dataset. These four events were all carried out by one group, and all of them record a consistent pre-pump phase in the hours before the pump. Fortunately, admins cannot use this technique regularly to avoid detection. Indeed, outsiders could detect this pattern to increase the probability of predicting the target coin. Moreover, these events often fail, and most users could lose trust in the admins and leave the group.

\noindent\textbf{Can pump and dump groups manipulate Bitcoin or major cryptocurrencies?} To answer this question, we make a short simulation. Let us take the buy volume on the first 10 minutes of the largest pump and dump we monitored. It is $31$ BTC on the SingularDTV (SNGLS). Now, we take a snapshot\footnote{Data retrieved on April 12, 2019} of the exchange order book for the trading pair BTC/USD. We assume that the market is frozen and only the pump and dump group members can take action. This is the best case for raising the price. We find that the amount of money at their disposal can increase the BTC value by less than \$5, which is way smaller than the natural oscillations of the coin in 10 minutes. So, the answer is no. Though these groups are very large, they cannot attack coins with large volumes like Bitcoin.

\noindent\textbf{Is it possible for the exchange markets to stop pump and dump schemes?} In this work, we show that it is possible to detect a pump and dump scheme as soon as it starts. We also believe that exchanges can catch better than us when a fraudulent scheme is in action. In fact, the data owned by the exchange is more fine-grained: It has complete knowledge of the kind of operations performed, their amount, and precisely who performed them during the scheme. Moreover, we notice that little policy enforcement can reduce the number of these market manipulations. As discussed before, on November 25, 2017, the BitTrex exchange announced that it actively discourages any market manipulation and will begin to punish the participants~\cite{bittrex}. Since then, the amount of pump and dumps in the exchange drastically decreased. We counted, before the statement, more than 50 pump and dumps in the five months from July to the end of November 2017, and only 48 events in more than three years after the statement.
Another countermeasure could be stopping transactions on a cryptocurrency when it gains or loses more than some threshold or giving special protection to cryptocurrencies with extremely low market capitalization and trading volumes. Moreover, some exchanges list cryptocurrencies with shallow trading volumes. De-listing these cryptocurrencies, as some exchanges do~\cite{cobinhood}, could make smaller groups desist.

\section{Conclusion}

In this work, we conducted an in-depth analysis of the pump and dump ecosystem. 
We studied the relationship between the groups, the exchange, and the target cryptocurrencies in a longitudinal analysis that spans over three years.
We thoroughly investigated the Big Pump Signal group and the pump and dump operations carried out by the Yobit exchange.
Moreover, we introduced our classifier that leverages the \textit{rush orders}, a peculiar kind of order that is particularly effective for the detection. The proposed classifier outperforms the state of the art both on performance (98.2\% of precision and 91.2\% recall against 52.1\% precision and 78.8\% recall) and on speed, moving the expected detection time from 1 hour to 25 seconds.
Given the lack of a pump and dump dataset, as a further contribution, we release to the community our resource~\cite{pumpdataset} of more than 900 confirmed pump and dumps to enable further studies on the topic.
precision and 78.8\% recall) and on speed, moving the expected time of detection from 1 hour to 25 seconds.
Finally, we moved on the crowd pump. Here, we conducted the first analysis based on data on this kind of operation and show the potential of a purely market-based approach to detect these kinds of events. 
We achieved promising results, catching the start of the two operations a few minutes after their opening and detecting the abnormal market conditions driven by tweets of celebrities.
We think that this work helps to understand a complex phenomenon, improves the awareness of the investors interested in the cryptocurrency market, and can help the authorities regulate this particular market in the future.

\section{Future work}
A possible future direction for this work is to integrate inside our system new features based on information extracted from social networks like Reddit or Twitter.
As seen before, Figure~\ref{fig:doge} and Figure~\ref{fig:xrp} show some correlation between Reddit submissions and the increase in the price of the coin. This suggests that information extracted from social networks may be used as a feature to identify crowd pumps. In particular, the advantage of integrating social media in the model may be crucial to help disambiguate cases where the rise in the price of the coin is not due to market manipulation but to solid market fundamentals.
During our study, we found a large number of signal groups. These groups are more significant than the pump and dump groups and arrange operations more frequently. As future work, it would be interesting to study the impact of these groups and their activity on the market.
We tested our classifier on two crowd pumps on the XRP and DOGE cryptocurrencies. It would be interesting to collect a larger dataset to further assess the performance of our detector.
Finally, it would be also interesting to verify if the methodology we developed to detect pump and dumps on the cryptocurrency market can also be used to detect these market manipulation in the stock market.

\section*{Acknowledgments}
This work was supported in part by the MIUR under grant ``Dipartimenti di eccellenza 2018-2022" of the Department of Computer Science of Sapienza University.

\newpage
\appendix
\section*{Appendix}
\setcounter{section}{1}
\subsection{Monitored Groups}
\label{ssec:monitoredgroups}
\begin{table}[h]
\captionsetup{width=0.7\textwidth}
    \caption{%
        This table reports the acronym that we will use in the charts, the extended name, and the Telegram link of each monitored group.
    }\label{tab:monitored_groups}
    \begin{tabular}{l l l}
        \toprule
        Group code &  Group name & Telegram link \\
        \midrule
        TCC & Trading Crypto Coach & https://t.me/tradingcryptocoach \\
        TCG & Trading Crypto Guide & https://t.me/TCGFORYOU \\
        BPS & BigPumpSignal & https://t.me/bigpumpsignal \\
        BPG & BigPumpGroup.com & https://t.me/bigpumpgroup\_com \\
        MPG & Trading Mega Pump Group &  https://t.me/mega\_pump\_group\\
        C4P & Crypto4Pumps & https://t.me/Crypto4Pumps \\
        PKG & Pump King Community & https://t.me/pumpingking \\
        ATW & AltTheWay & https://t.me/AltTheWay \\
        LUX & Luxurious Crypto & https://t.me/LuxuriousCrypto \\
        CPS & Cryptopia pump squad & https://t.me/cryptoflashsignals \\
        CCB & Crypto coin B & https://t.me/CryptoCoinsCoach \\
        FCS & Fast Crypto Signals & https://t.me/fastcrypt \\
        WCG & Whales Crypto Guide & https://t.me/Whalesguide \\
        TWP & Today We Push & https://t.me/TodayWePush \\
        CP & CrypticPumps & https://t.me/CrypticPumps \\
        BPF & Big Binance Pump Family & https://t.me/rocketpumptrader \\
        CW & Crypto Waves & https://t.me/CryptoCoinsWaves \\
        CCS & Coin Coach Signal & https://t.me/CoinCoachSignals \\
        SE & Signal Express & https://t.me/signalexpresss \\
        CPI & Crypto Pump Island & https://t.me/crypto\_pump\_island \\

        \bottomrule
    \end{tabular}
    
\end{table}

\newpage
%\section*{Appendix}
\bibliographystyle{ACM-Reference-Format}
\bibliography{biblio}

\end{document}